\documentclass[journal]{IEEEtran} 
\usepackage[utf8]{inputenc}
\usepackage{amsmath,color,graphicx,amssymb,mathtools}
\usepackage{ifthen}
\usepackage{textcomp}
\usepackage{gensymb}
\usepackage{placeins}
\usepackage{float}
\usepackage[bookmarks=false]{hyperref}
\usepackage{tabularx}
\usepackage{epstopdf,multirow}
\usepackage{xcolor}
\usepackage{mathrsfs}

\usepackage{tikz,tikz-3dplot}
\usetikzlibrary{patterns}
\usepackage{cite}
\usepackage[T1]{fontenc}

\usepackage{balance}

\newcommand{\rhovec}{\boldsymbol{\rho}}
\newcommand{\rhodvec}{\dot{\rhovec}}
\newcommand{\rhoddvec}{\ddot{\rhovec}}

\begin{document}
\bstctlcite{IEEEexample:BSTcontrol}
\pagenumbering{gobble}

\title{Space Surveillance with High-Frequency Radar}
\author{\IEEEauthorblockN{Brendan~Hennessy, Heath~Yardley, Rob~Debnam, Tristan~A.~Camilleri, Nicholas~K.~Spencer, David~A.~Holdsworth, Goeff~Warne, Brian~Cheung, Sergey~Kharabash
}\\
\IEEEauthorblockA{Defence Science and Technology Group, Edinburgh, SA 5111, Australia.}\\
}

\maketitle
\begin{abstract}
High-Frequency (HF) radar is well suited to the surveillance of low-Earth-orbit space. For large targets, a small deployable HF radar is able to match the detection performance of much larger space surveillance radar systems operating at higher frequencies. However, there are some unique challenges associated with the use of HF, including the range--Doppler coupling bias, coarse detection-level localisation, and the presence of meteor returns and other unwanted signals. This paper details the use of HF radar for space surveillance, including signal processing and radar product formation, tracking, ionospheric correction, and orbit determination. It is shown that by fusing measurements from multiple passes, accurate orbital estimates can be obtained. Included are results from recent SpaceFest trials of the Defence Science and Technology Group's HF space surveillance radar, achieving real-time wide-area surveillance in tracking, orbit determination, and cueing of other space surveillance sensors
\end{abstract}


\section{Introduction}

\IEEEPARstart{E}{arly radar} systems were developed at HF, typically defined as the range 3--30 MHz, due to the limitations of the available hardware at the time. As technology improved, most radar systems were developed at higher frequencies to utilise the benefits offered, with HF being mainly used to exploit unique propagation properties, notably skywave over-the-horizon (OTH) propagation via the ionosphere~\cite{cameron1995jindalee,9764214} and surface-wave propagation via diffraction~\cite{anderson2003investigations}. However, there has still been considerable work undertaken in HF radar used in a \textit{standard} line-of-sight (LOS) mode. LOS HF radar is beneficial, or maybe even an ideal option, for several areas of research, with examples including meteor studies~\cite{holdsworth2004buckland}, upper atmosphere wind profiling~\cite{singer2008new}, ionospheric research~\cite{chisham2007decade}, missile defence~\cite{8835810}, and space surveillance.

The first operational surveillance radars were HF, most notably Chain Home~\cite{neale1985ch}, although Germany also developed HF radar systems~\cite{6470438}, including a passive radar     system~\cite{griffiths2010klein}. Following the advent of radar, most systems operated at higher frequencies. However, there has been significant research and work undertaken into the use of LOS HF radar. Specific areas that are relevant to this paper include the use of HF radar for monitoring rocket tests and launches~\cite{bowser1966radar,sinnott1988development,2016793,818176}. Additionally, the use of OTH radars (OTHRs), operated in a LOS mode, has been considered for detecting targets above the ionosphere, including satellites and other resident space objects (RSOs)~\cite{headrick1962madre, colbert2004analysis, 10501529,henault2025space}. More recently, LOS HF radars have been used directly for space surveillance purposes~\cite{6651980, frazer2013orbit, czarnowske2023space}. There has also been relevant work undertaken on space surveillance systems utilising frequencies near HF (such as low Very High Frequency (VHF))~\cite{holdsworth2020low, heading2024micro, holdsworth2024space}. LOS HF radars may also be used as part of hybrid systems, which incorporate LOS aspects such as forward-deployed receivers contributing to an HF radar network to receive reflections from skywave transmitters (referred to as SkyLOS)~\cite{frazer2007forward,10371074, 10548923, bok2024key}.

{Whilst the focus of this paper is on HF space surveillance radar, the vast majority of current space surveillance systems operate at higher frequencies. Some examples include the Space Surveillance Network, with radars operating at frequencies spanning from VHF through to C-band~}\cite{USSPACECOM}{, LeoLabs' radar network operating at ultra high frequency (UHF) and S-band}~\cite{rowland2021worldwide}, {the GRAVES continuous-wave (CW) bistatic system operating at VHF}~\cite{th2005graves}{, monostatic and bistatic systems at L-band~}\cite{7485184, 7832621,gomez2019initial}{, a cooperative bistatic radar system utilising a radio telescope receiver at low UHF}~\cite{cutajar2018real,9590991}{, non-cooperative bistatic radar systems utilising radio telescope receivers at VHF~}\cite{tingay2013detection, 7944483, malanowski2021passive},{~and a dedicated passive radar space surveillance observatory at VHF~}\cite{finch2024passive}{. These systems typically only surveil a limited volume of space, either through a single beam or single fence, for a monostatic radar, or the intersection of beams for a bistatic radar. These narrow-beam surveillance radars are able to achieve wide-area space surveillance by rapidly scanning the entire field of regard in conjunction with a very high-powered transmitter, although such systems are very costly and may take years to deploy. A detailed comparison between these types of operational radars and an HF radar is contained here~}\cite{8835810}{. Some of the other bistatic systems are able to surveil a large volume of space by relying on significant stand-off distances between the transmitter and the surveillance volume of interest, although this may limit the system sensitivity.

HF radar is well suited for wide-area space surveillance, with deployable systems being able to match the detection performance of much larger, more traditional space surveillance radar facilities. Long wavelengths allow instantaneous floodlight illumination of a wide volume and also a large antenna capture area to be achieved with a relatively small number of antennas and channels. By way of example, the results included in this paper all came from systems with transmitter beamwidths of 90\textdegree.}  However, there are some HF-specific challenges that need to be overcome, such as coarse detection-level localisation, errors introduced by the ionosphere, and the presence of clutter and other signals. A goal of this paper is to highlight some of the orbital-specific signal processing methods, as well as the approach to orbit determination, to overcome and mitigate some of these drawbacks to achieve sufficient trajectory estimation for persistent space surveillance.

This paper illustrates these approaches and methods with results from deployments of the Defence Science and Technology Group's (DSTG) High-Frequency Line-of-Sight (HFLOS) radar to recent SpaceFest activities in the Far North region of South Australia. SpaceFest is a space surveillance capability demonstration and equipment trial conducted by DSTG, which has occurred regularly since 2018. The HFLOS radar is an experimental HF wide-area surveillance staring radar~\cite{6651980}, enabling rapid prototyping and evaluation. It is an excellent development testbed for HF technology, algorithms, and processes~\cite{7485242,10371125,10371098,10371197,jonker2024doppler,hflos_iono_corrections_tristan}. These deployments have allowed for continuous upgrades and evaluations of the system, and enabled the HFLOS radar to be deployed in a variety of configurations, including different receive array layouts, multiple sites for quasi-monostatic, bistatic, and multistatic operation, and being co-deployed with other sensors, all alongside general system-level and processing improvements.  The HFLOS radar system is able to perform real-time and persistent wide-area space surveillance using long-coherent integration time (CIT) processing and a short inter-CIT stride time.

The layout of this paper is as follows. Section \ref{sec:sec2} provides a brief overview of an HF radar system, including specific aspects of a space surveillance LOS system. Section \ref{sec:sec3} details all the signal processing steps in forming the radar products, including the range--Doppler maps, Doppler-walk/smear mitigation, detection parameters, and tracking. Section \ref{sec:sec4} covers the impact of the ionosphere on the HF radar measurements as well as means of reducing these impacts with the use of an ionospheric model. Section \ref{sec:sec5} discusses the space surveillance deployments of the HFLOS radar at the SpaceFest trials and some of the various configurations. Section \ref{sec:sec6} details the orbit determination (OD) process, including the initial orbit determination (IOD) steps, multi-pass OD, and the challenges of determining orbits from detections with linear arrays. This paper concludes with Section \ref{sec:sec7}. An Appendix \ref{sec:appendix_telescope} is included to detail DSTG's electro-optical (EO) telescope space surveillance systems, used in joint space surveillance with the HFLOS radar.

\section{System Description}\label{sec:sec2}

{Results included in this paper are drawn from a series of deployments of the HFLOS radar from 2018 to 2022. Every aspect of the HFLOS system has been undergoing continuous improvement for many years, and so each deployment was unique. The specific focus of this paper is on the signal processing, OD, and space surveillance; however, these are only possible as a result of significant engineering and software development. 
Only a few aspects of the many subsystems which comprise an HF system will be covered in the following.}

\begin{figure}[ht]
\begin{center}
\includegraphics[width=\columnwidth]{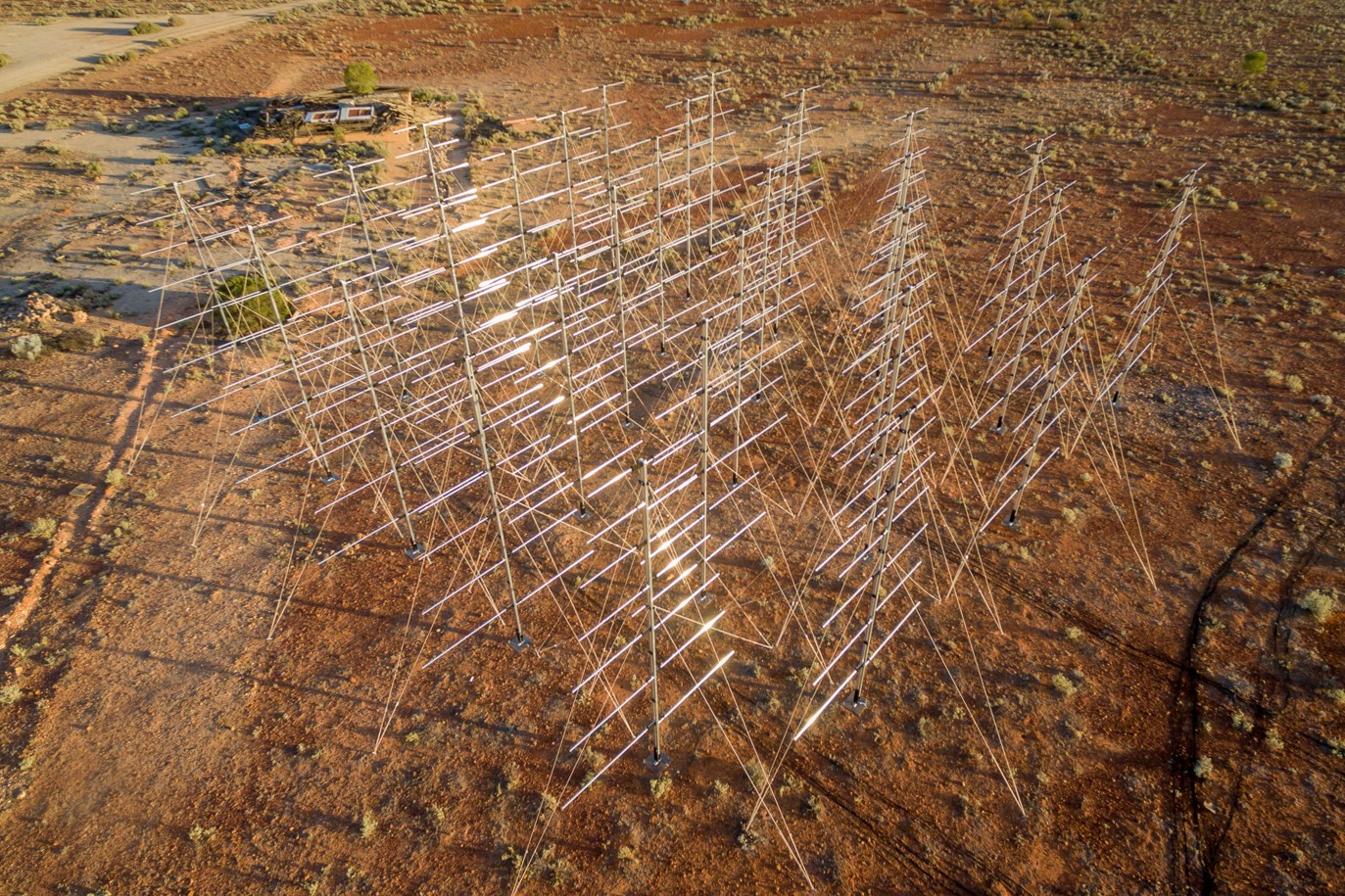}
\end{center}
\caption{Wide-view photograph of a compact planar receive array, directed toward zenith.}
\label{fig:hflos_rx}
\end{figure} 

\begin{figure}[ht]
\begin{center}
\includegraphics[width=\columnwidth]{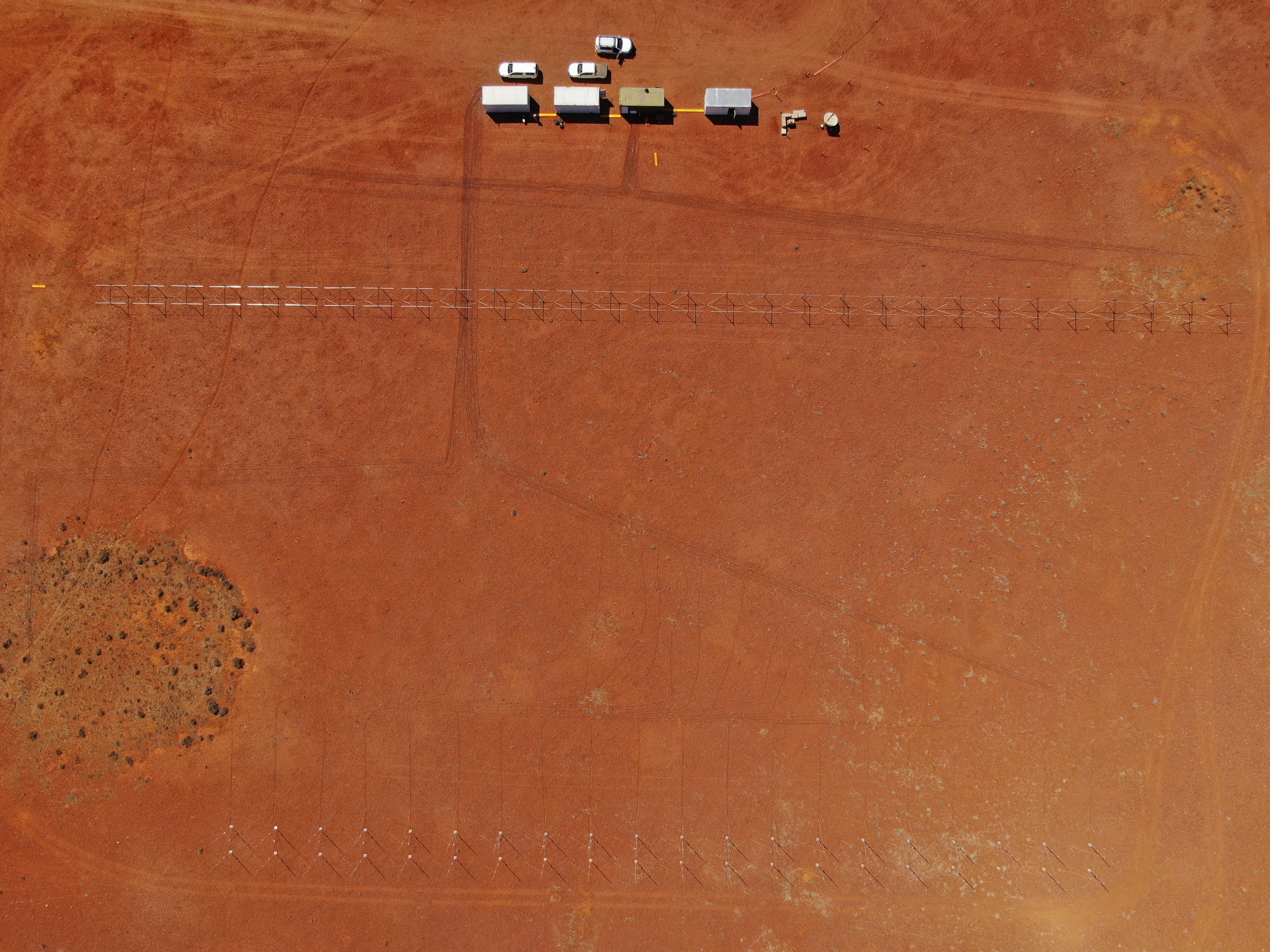}
\end{center}
\caption{Wide-view photograph of two linear receive arrays, each array element consists of two receive antennas. Not pictured, to the left was a third linear array, perpendicular to the pictured arrays.}
\label{fig:hflos_rx2}
\end{figure}

The HFLOS transmit system consists of multiple elements, with the total transmit power arising from freespace combination from a transmit array. This allows for a modular design and deployment of the waveform generators, power amplifiers, and transmit antenna for each transmit channel. HF wavelengths ensure a large beam for wide-area illumination. This coverage can be broadened with physical pointing or phase adjusting (beam-spreading) the transmit array to form an instantaneous wide field of regard.

{Although the waveform generators allow the use of arbitrary signals, linear-frequency-modulated (LFM) \textit{chirps} are typically used as part of an LFMCW waveform. An LFMCW ensures the amount of power on a target is maximised, and improves range resolution, whilst assisting to discriminate against noise and other spurious signals~}\cite{klauder1960theory}. The main drawbacks of this waveform class are the ambiguous range and Doppler measurements, as well as the impact of the range--Doppler coupling bias~\cite{fitzgerald1974effects}. These drawbacks will be addressed in greater detail in Section \ref{sec:detection_params}.

Only the upper end of the HF spectrum is used for space surveillance purposes, as the signals need to penetrate the ionosphere. A typical system would operate with a centre frequency in the range of  28--32 MHz. This higher frequency also benefits the detection of smaller RSOs, as the Rayleigh scattering results in a significantly reduced radar cross-section (RCS) at lower frequencies~\cite{6651980}. It is for this reason that HF radar will typically only observe larger, albeit more defence-relevant, RSOs, and will not detect cubesats and small debris.

{Even at the upper end of the HF spectrum, skywave propagation modes may still be supported at low grazing angles, and so there will be significant clutter and interference from environmental and human-made noise, as well as signals from other radar and communication systems due to spectrum congestion. Because of these unwanted signals, and the noise figure of typical receivers, HF radar receive systems are almost always externally noise-limited~}\cite{6651980}{.}

Previous system deployments have included quasi-monostatic operation, where the transmitter and receiver are separated by several kilometers (required in order to reduce the direct wave power observed by the receiver), as well as being deployed in bistatic and multistatic configurations, where the transmitter and one of the receiver systems are separated by several hundred kilometers. 
    
A myriad of receive array configurations have also been utilised, from linear arrays, compact planar, and sparse circular configurations to multistatic combinations of these receive arrays. {The receive array configuration determines the spatial accuracy of the radar and which spatial dimension the radar can discriminate, as well as the compromise between accuracy and other aspects such as sidelobes and grating lobes; this is detailed further in Section }\ref{sec:sec5}{.} An example planar receive array configuration is shown in Figure \ref{fig:hflos_rx}, consisting of 30 log-periodic dipole array (LPDA) antennas directed toward the zenith. An example of a multistatic linear array configuration is shown in Figure \ref{fig:hflos_rx2}.

\section{Radar Product Formation}\label{sec:sec3}

There are significant advantages in the use of HF radar for space surveillance, allowing modest-sized or deployable HF radars to match the detection performance of much larger systems for large targets. These benefits have been detailed previously~\cite{8835810}, but will be broadly summarised here.

The low frequencies and longer wavelengths mean a large antenna capture area can be covered by a relatively small number of receiver channels. These large wavelengths also readily enable broad \textit{floodlight} illumination of a wide volume. {The low bandwidths of HF radar, typically less than 30 kHz}, in conjunction with pulse repetition, mean there will only be a very small \textit{nucleus} of pulses, after range compression, to cover all velocities and ranges. There are drawbacks to the ambiguous range and Doppler parameters (covered in greater detail below), but a great advantage is the very small radar data products. For a CIT of $T$ seconds, and a waveform repetition frequency (WRF) of $f_p$~Hz, the number of pulses is $Tf_p$, typically chosen to be a power of two, enabling efficient Discrete Fourier Transform implementation. For a pulsed radar with a bandwidth of $B$ Hz, the total number of cells in a delay--Doppler map is $TB$. When combined with the relatively few receive channels, these small radar data products readily allow longer CITs, short inter-CIT stride times, many receive beams, and advanced signal processing algorithms. 

Significant recent improvement has been driven by increased computational power, combined with efficient {graphics processing unit (GPU)} implementations of algorithms. Compared to an earlier central processing unit (CPU) baseline operating a one~second CIT, and 20 beams~\cite{6651980}, the system is now capable of performing real-time wide-area surveillance with a 10.24~s CIT, 0.5~s stride time, many hundreds of receive beams, and hundreds of Doppler-rate hypotheses.

There are some specific drawbacks to the use of HF radar. The use of such large wavelengths with low bandwidths means any individual detection's geolocation will not be as precise as that of higher-frequency systems, due to the large beamwidth and coarse range resolution. Doppler, on the other hand, will be very precise, due to the long CITs. The detection level inaccuracy is compounded by the impact of the range--Doppler coupling bias, resulting from the (relatively) low chirp rates alongside the high velocities of RSOs~\cite{klauder1960theory,fitzgerald1974effects}. The low frequencies also mean the impact of significant clutter from ionised meteor trails, skywave backscatter, and other interference. The presence of this clutter and interference is another reason why LFM waveforms are preferred over waveforms with unambiguous detection parameters, due to the limitations that would be introduced by the presence of a pedestal floor in the ambiguity function. Finally, the low frequencies also result in detection-level errors due to the refraction of the ionosphere. Many of these drawbacks can be mitigated or overcome with downstream processing, such as aspects in the signal processing, tracking, and orbit determination.

\subsection{Range--Doppler Map Formation}
Long-CIT processing is important in order to improve sensitivity, as most RSOs will be small enough that Rayleigh scattering significantly reduces the RCS. The standard pulse-Doppler approach, referred to here as `match at zero Doppler', is \mbox{straightforward~\cite{klauder1960theory,1163621,stove1992linear}}. Range-compressed pulses are combined through pulse integration, essentially with the Discrete Fourier Transform across slow time, to resolve Doppler. This is very fast computationally, and will allow efficient implementation of short stride times.

\begin{figure}[ht]
\begin{center}
\includegraphics[width=\columnwidth]{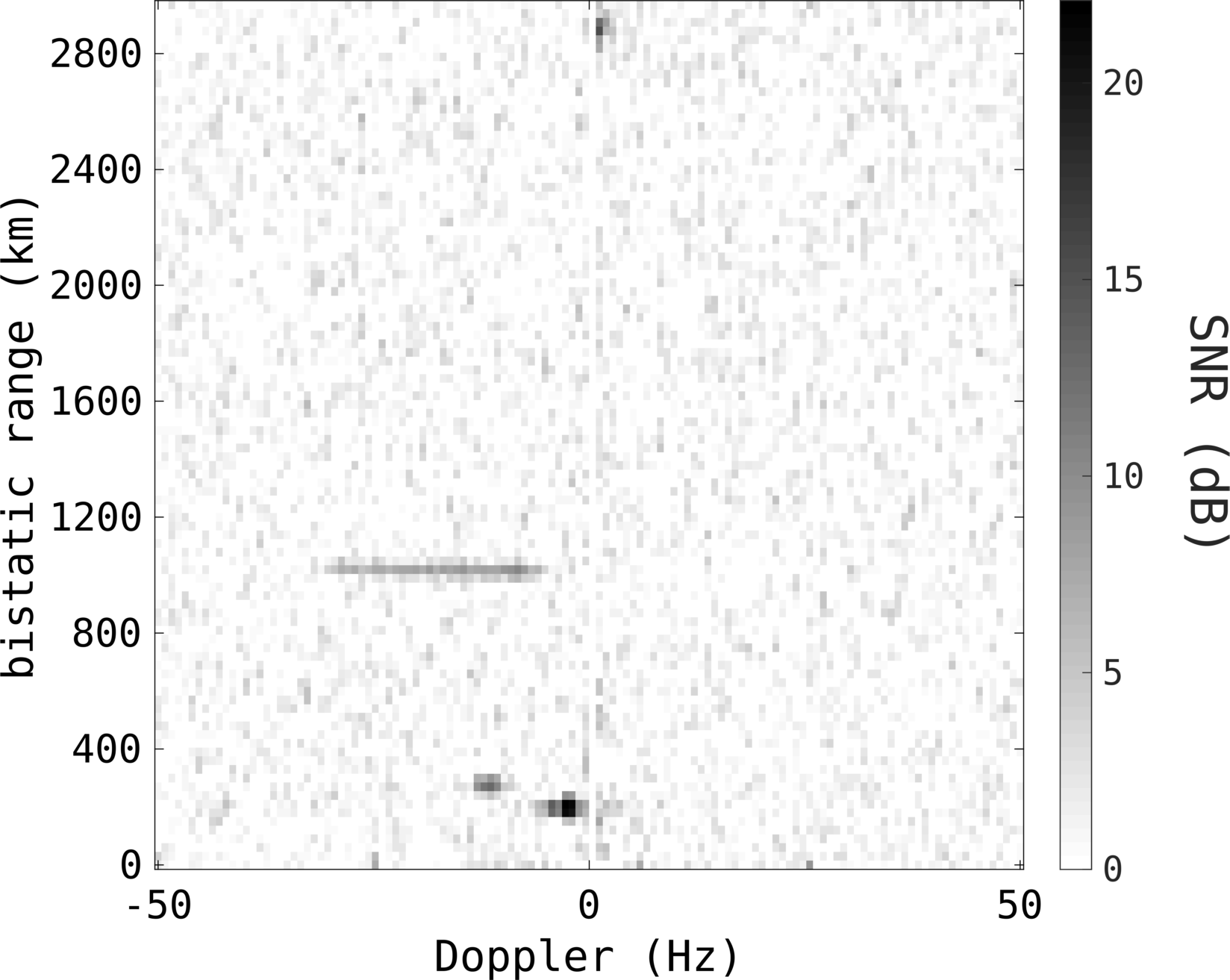}
\end{center}
\caption{An example range--Doppler map formed by the HFLOS radar, using a CIT of 1.28~s.}
\label{fig:typical_ddmap_space}
\end{figure}

Figure \ref{fig:typical_ddmap_space} is a typical example of an HFLOS radar range--Doppler map, formed with a CIT of 1.28~s. Apparent in the range--Doppler map is a satellite moving through Doppler at an approximately 1000~km range~(the satellite's Doppler spans approximately 20 Hz). Also present are some strong meteor returns at low range. The direct wave from the transmitter has been nulled by using a Doppler notch at 0~Hz. There are also some potential skywave returns at 2800~km.

As there are typically many more beams to be formed than the number of receivers, range--Doppler maps are formed for each antenna receive channel, and then these channel maps are beamformed to generate a range--Doppler map corresponding to a direction of arrival of interest. Detections are formed from each beam's range--Doppler map, with a constant false alarm rate (CFAR) detection scheme~\cite{turley1997hybrid}.

Apart from the significant propagation distances, the biggest challenge to achieving sufficient sensitivity with long-CIT processing is the migration, or smearing, of RSOs through the range--Doppler map. RSOs in low earth orbit (LEO) will be travelling at speeds in excess of 7 km/s, and so there will be considerable motion in every measurement aspect, for even modest CITs. Spatial migration can be mitigated and controlled with array size. Reducing the signal bandwidth, within reason, can mitigate range migration to an extent. However, the biggest challenge is Doppler migration. The extreme speeds result in a changing geometry over a long CIT, and the Doppler resolution increases with CIT length itself, and so the result is considerable Doppler-walk of RSO radar returns with any long-CIT radar system. This can have a significant impact on the signal-to-noise ratio (SNR), and so mitigating Doppler-walk is a primary focus.

\subsection{Doppler-Walk Mitigation}\label{ssec:doppler_walk}
The significant orbital speeds of any RSO, combined with the long CIT required for sufficient sensitivity, will result in considerable Doppler-walk. This Doppler-walk, or Doppler migration, will degrade the sensitivity as the radar returns are spread across a large number of Doppler cells, limiting the SNR in any given cell. The range--Doppler map example in Figure \ref{fig:typical_ddmap_space} illustrates considerable Doppler-walk, with the satellite spanning 20 Hz Doppler, despite the short 1.28~s CIT. As the smearing results from the changing relative radial velocity, the mitigation is often colloquially referred to as `acceleration processing'. However, this term is used because the slant-range acceleration is from the perspective of the radar, and not the target's true dynamics. Satellites and RSOs experience less than 1 g of acceleration and exhibit Doppler-walk, and even the changing geometry from constant-velocity motion can result in significant Doppler-walk.

Due to the Doppler-walk's deleterious impact on the SNR, any mitigation will result in significant sensitivity improvements in detecting RSOs. Doppler-rate processing has been a key facet of fast-target detection with radar for many years~\cite{kelly1961radar,4323176,1163621}. Some relevant examples for this paper are results showing sensitivity gains, as well as improved target discrimination, for missile tests observed with {OTHR~\cite{mcgeogh1967acceleration,Jensen1966madre,Jensen1966madre2},} 
 missile tests observed by LOS HF radars~\cite{818176,2016793}, and also LOS space surveillance radar with sufficiently long integration {times~\cite{8835821, hennessy2022deployable, jkedrzejewski2024exploring, klare2024future}.}  Interestingly, the ray-focusing effect of skywave propagation actually dampens the apparent Doppler-walk that results from fast-changing geometry; however, it is still useful for long-CIT processing~\cite{mcgeogh1967acceleration}.

Not only does this Doppler-rate processing drastically improve the sensitivity against high velocity and accelerating targets, it can also greatly assist in discriminating valid targets of interest from clutter, meteor effects, interference, and other slow-moving targets.

The approach to Doppler-rate processing for the HFLOS system is to apply an inverse Discrete Fourier Transform to the Doppler from each beam's range--Doppler map to generate the slow-time pulse stack. A slow-time chirp is applied at each range bin to mitigate the motion-induced chirp. Potential chirp rates are searched through and any that result in a sensitivity improvement, by focusing the Doppler signal, are chosen as valid detections. Care is taken to attempt to ensure that only valid chirp rates are selected, that is, chirps that focus returns for a sensitivity gain and not simply smearing clutter returns throughout Doppler.

\begin{figure}[ht]
\begin{center}
\includegraphics[width=\columnwidth]{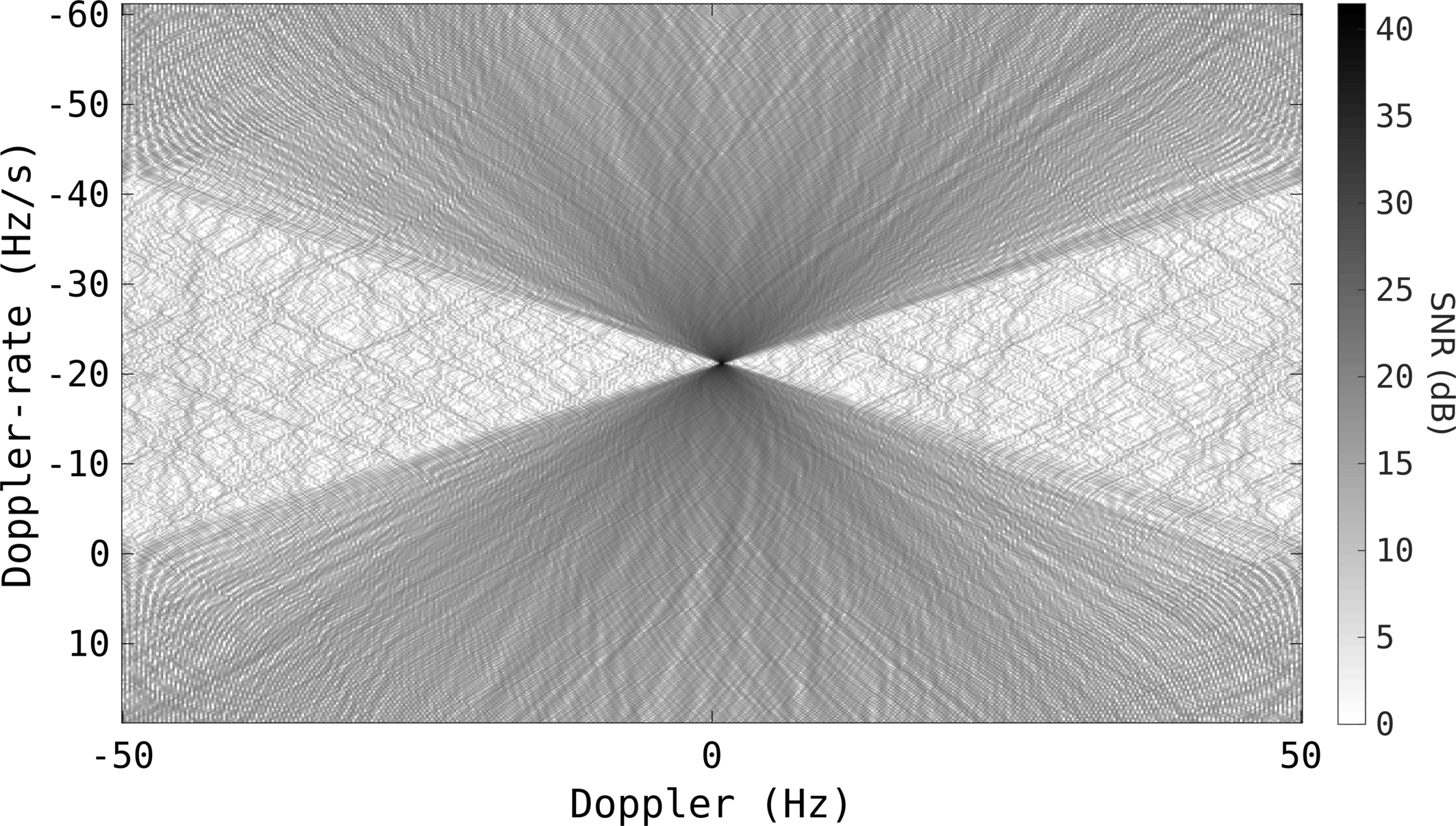}
\end{center}
\caption{{Chirpogram} of a single range bin showing the SNR of the Doppler signal for each different Doppler-rate hypotheses, with a CIT of 5.12~s. The RSO is the same one as from Figure \ref{fig:typical_ddmap_space}.}
\label{fig:classic_doppler_bowtie}
\end{figure}

Figure \ref{fig:classic_doppler_bowtie} is the \textit{chirpogram} of the radar returns from the same satellite as in \mbox{Figure \ref{fig:typical_ddmap_space}.} The slow-time returns from the satellite's range bin have been processed with a large span of Doppler-rate hypotheses. A 5.12~s CIT was used so that the satellite spans the whole Doppler extent when uncompensated, i.e., Doppler rate = 0 Hz/s. At the \textit{correct} Doppler rate, where the SNR is maximised, the Doppler-walk is completely mitigated and the slow-time signal is focused to a single Doppler bin, providing a significant increase in sensitivity as well as better localising the target.

\begin{figure}[ht]
\begin{center}
\includegraphics[width=\columnwidth]{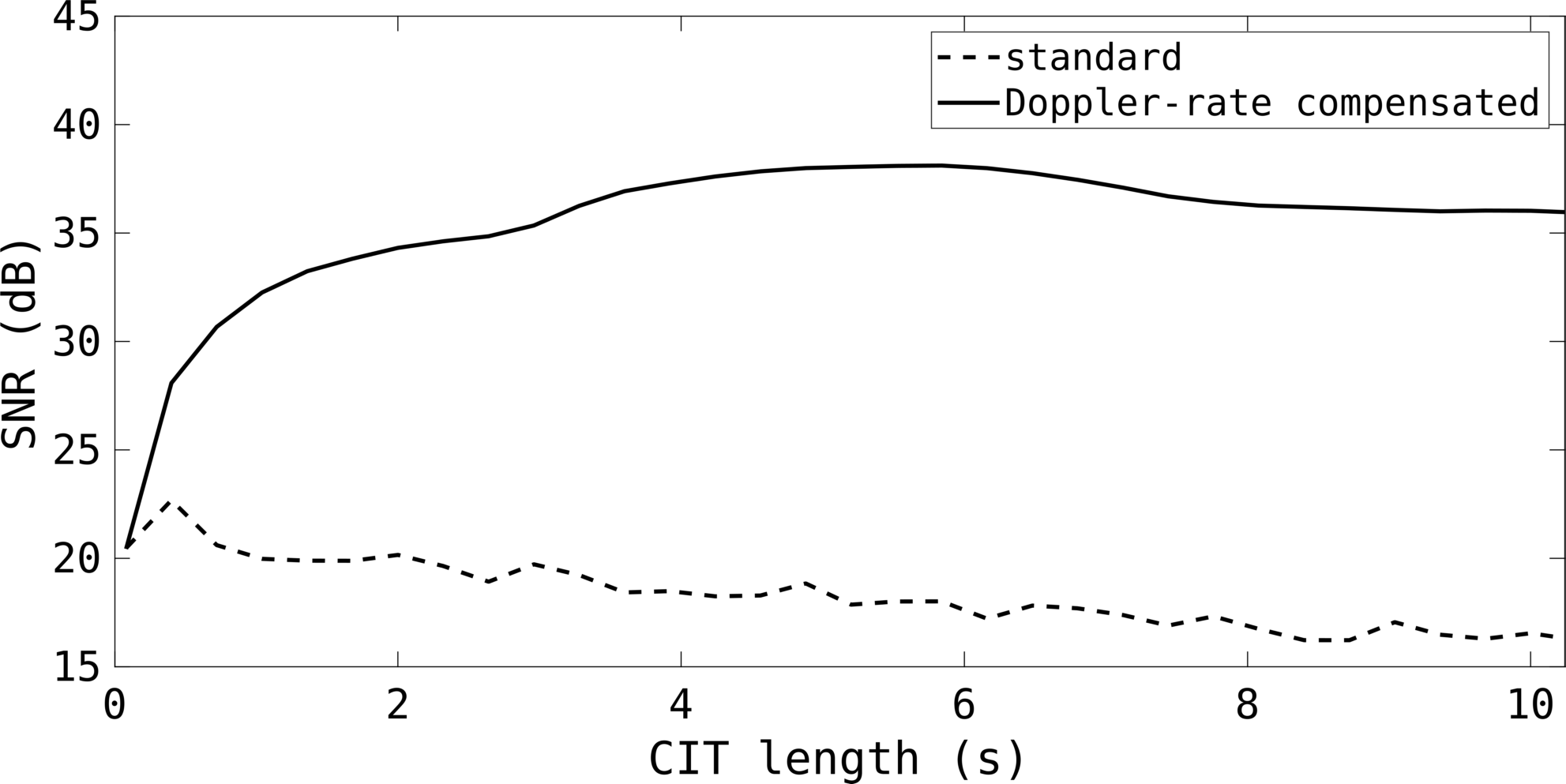}
\end{center}
\caption{The SNR of an example RSO against CIT length for standard processing as well as Doppler-walk mitigation.}
\label{fig:SUPER_accel_snr}
\end{figure}

As illustrated in Figure \ref{fig:SUPER_accel_snr}, the SNR improvement with Doppler-rate processing is evident. Even with only modest CIT lengths, there is a significant increase in sensitivity. The specific shape of the sensitivity curve will be different for every target, depending on the motion and the range. The limiting factor may well be range migration, or even potentially angular migration, at long CITs. Note that this target had a very significant RCS, as it was able to be detected with a CIT less than 1~s. With the HFLOS system, most RSOs will not be detectable without Doppler-rate compensation processing, regardless of the CIT length.

\begin{figure}[ht]
\begin{center}
\includegraphics[width=\columnwidth]{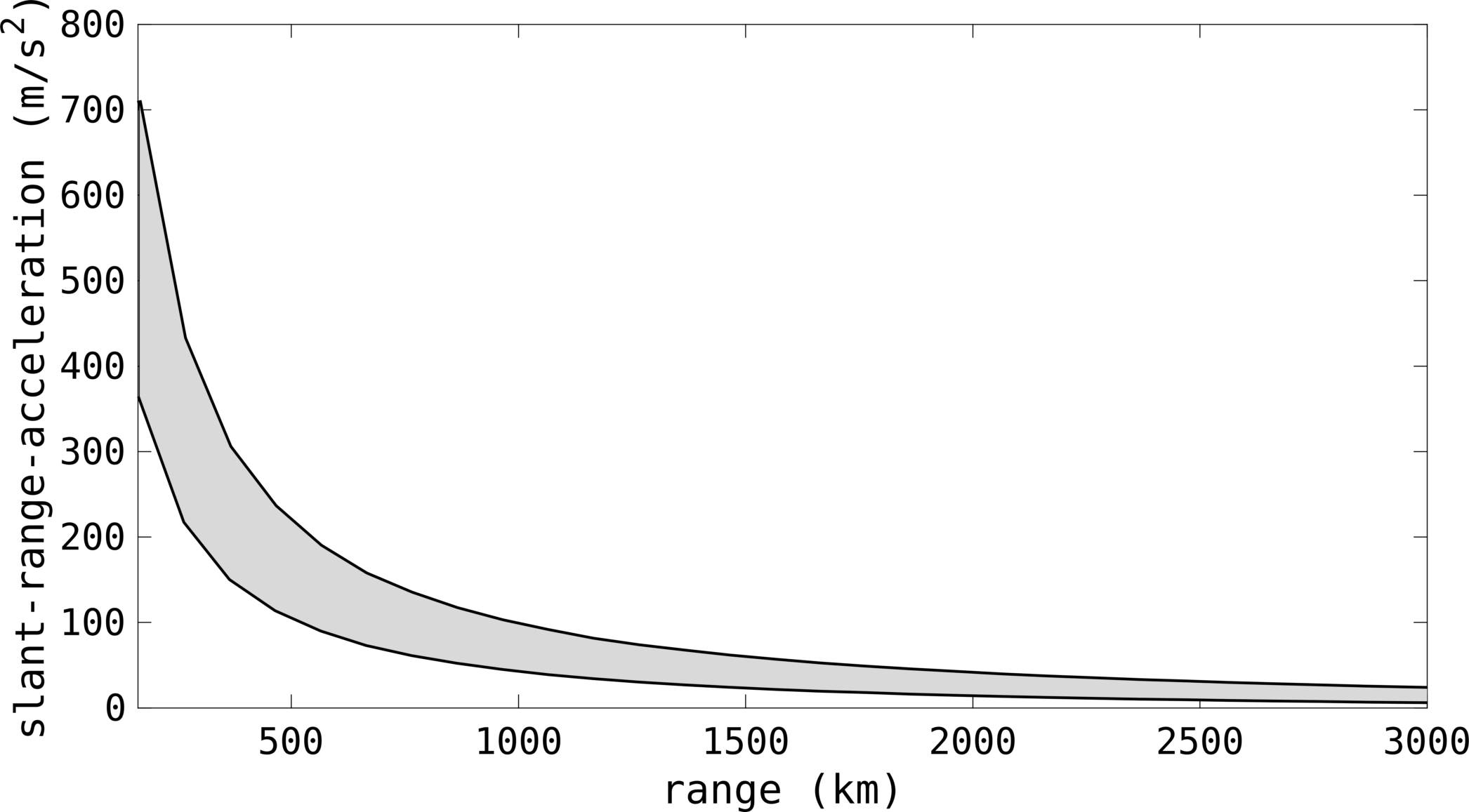}
\end{center}
\caption{The span of valid slant-range acceleration values for an object in an Earth-centered orbit, for a zenith beam from a monostatic radar.}
\label{fig:slant_range_acceleration_rate}
\end{figure}

For all potential RSOs, there is an incredibly large overall span of Doppler rates that would need to be searched. However, for a single point in space, for an Earth-centered orbit, this span is considerably reduced~\cite{8448187}. Figure \ref{fig:slant_range_acceleration_rate} shows the span of valid slant-range accelerations for a zenith beam from a monostatic radar. Although it spans from near-zero through to over 700~m/$\text{s}^\text{2}$, for a specific range bin the span is quite limited. This can improve processing by reducing the number of Doppler-rate hypotheses to only match for valid orbits. Prior to operation, a radar could determine the valid Doppler-rate span for each beam to reduce processing load whilst still matching every potential Earth-centered orbit.

A larger separation between the transmitter and the receiver will greatly lower the extent of any potential bistatic slant-range acceleration, as the transmitter--target--receiver geometry does not change as quickly with a larger transmitter--receiver baseline.

\subsection{Detection parameters}\label{sec:detection_params}

 Given the positional vector (slant-range vector) between the radar receiver and the target, $\rhovec_{r}$, as well as its time derivatives $\rhodvec_{r}$ and $\rhoddvec_{r}$, the receiver slant range, slant-range rate, and slant-range acceleration are given by
\begin{IEEEeqnarray}{rCL}
  \rho_{r} &=& \lvert \rhovec_{r}\rvert~,\label{eq:slant_range}\\
  \dot{\rho}_{r} &=& \frac{\rhovec_{r}\cdot\rhodvec_{r}}{\rho_{r}}~, \\
  \ddot{\rho}_{r} &= -&\frac{(\rhovec_{r}\cdot\rhodvec_{r})^2}{{\rho_{r}}^3} +  \frac{|\rhodvec_{r}|^2 + \rhovec_{r}\cdot\rhoddvec_{r}}{\rho_{r}}~,
\end{IEEEeqnarray}
and similarly for the transmitter terms $\rhovec_{t}$, $\rhodvec_{t}$, $\rhoddvec_{t}$, $\rho_{t}$, $\dot{\rho}_{t}$, $\ddot{\rho}_{t}$.

{For a LFMCW radar with a centre frequency of $f$, and a WRF of $f_p$, transmitting a chirp waveform consisting of a bandwidth of $B$} with a chirprate of $\gamma = f_pB$ (the sign of $\gamma$ specifying the direction, or polarity, of the chirp), the measured bistatic range (the round-trip time delay multiplied by the speed of light, $c$), bistatic Doppler and bistatic Doppler-rate are given by:
\begin{IEEEeqnarray}{rCL}
    r_{m}& = & \rho_{r} + \rho_{t} + \frac{f}{\gamma}\left(\dot{\rho}_{r} + \dot{\rho}_{t} \right)+J\frac{c}{f_p}~,\label{eq:measured_bistatic_range}\\
    f_{d} & = &-\frac{f}{c}\left(\dot{\rho}_{r} + \dot{\rho}_{t} \right) + Kf_p~,\\
    \dot{f}_d & =& -\frac{f}{c}\left(\ddot{\rho}_{r} + \ddot{\rho}_{t} \right)\label{eq:doppler_rate_hectic}~,
\end{IEEEeqnarray} where $J$ is the number of times a target has spanned, or wrapped, in the unambiguous range, and $K$ is the number of times the target has spanned the unambiguous Doppler extent, sometimes referred to as the fold factor, such that
\begin{IEEEeqnarray}{rCL}
    0 &\leq r_m  &< \frac{c}{f_p}~,\\
    -\frac{f_p}{2} &\leq f_d  &< \frac{f_p}{2} ~.
\end{IEEEeqnarray}

The WRF is typically chosen to ensure a suitably large unambiguous range, in the order of a thousand kilometres, so that only very large RSOs in medium Earth orbit (MEO) are detected at ambiguous ranges. Care needs to be taken to ensure the chosen WRF does not place meteor clutter and other unwanted returns in the same range as the targets.  Also, for bistatic radar, the range measurement includes the baseline distance between the transmitter and the receiver, which needs to taken into account. For a monostatic radar, or quasi-monostatic radar, the measured bistatic range above is halved to estimate the target range directly.

Most radar systems are able to ignore the impact the use of LFM waveforms has on the measurement parameters. However, this is not the case for HF radar, especially with the significant orbital velocities of RSOs. The measured time delay is shifted by the Doppler frequency divided by the chirp rate, $\gamma$, due to the correlation in the matched-filter processing; this bias is referred to as range--Doppler coupling~\cite{fitzgerald1974effects}. The impact of range--Doppler coupling results in the offending term $\frac{f}{\gamma}\left(\dot{\rho}_{r} + \dot{\rho}_{t} \right)$ in \eqref{eq:measured_bistatic_range}. Interestingly, it can be shown that the range--Doppler coupling impact on the measured time delay, usually given by $\frac{f_d}{\gamma}$, can be more accurately expressed as $\frac{f_d}{\gamma - \frac{1}{2}\dot{f}_d}$~\cite{howard_rdrr_coupling}. However, this will have little impact on the actual delay, as any reasonable chirp rate will be significantly larger than any potential Doppler rate.

Because the Doppler measurement is ambiguous, the range--Doppler coupling bias can be challenging to remove. It is possible, however, to compare the Doppler once it has been unwrapped with the estimated Doppler from the rate of change of the measured range to determine the correct offset. Because of the coarse range resolution, a longer track is required to better estimate the measured range rate. With a sufficiently long track and an unwrapped Doppler ${f_{d}}'$ (if unwrapped correctly, there will be one single correct constant offset $K$ for the whole track), the estimated rate of change of the measured range, $\dot{r}_m$, can be be used to estimate the fold factor $K$:

\begin{IEEEeqnarray}{rCL}
    \hat{K} = \text{round}\left[\frac{1}{f_p}\left(  {f_{d}}' + \frac{f}{c}\dot{r}_m + \frac{f}{\gamma}\dot{f}_d \right)\right]~.\label{eq:fold_factor_estimate}
\end{IEEEeqnarray}  With the fold factor estimated, the true Doppler can then be used to correct for range--Doppler coupling and determine an unbiased range estimate:
\begin{IEEEeqnarray}{rCL}
    \hat{r}_m = r_m + \frac{c}{\gamma}\left({f_{d}}' - \hat{K}f_p \right)~.
\end{IEEEeqnarray}

\begin{figure}[ht]
\begin{center}
\includegraphics[width=\columnwidth]{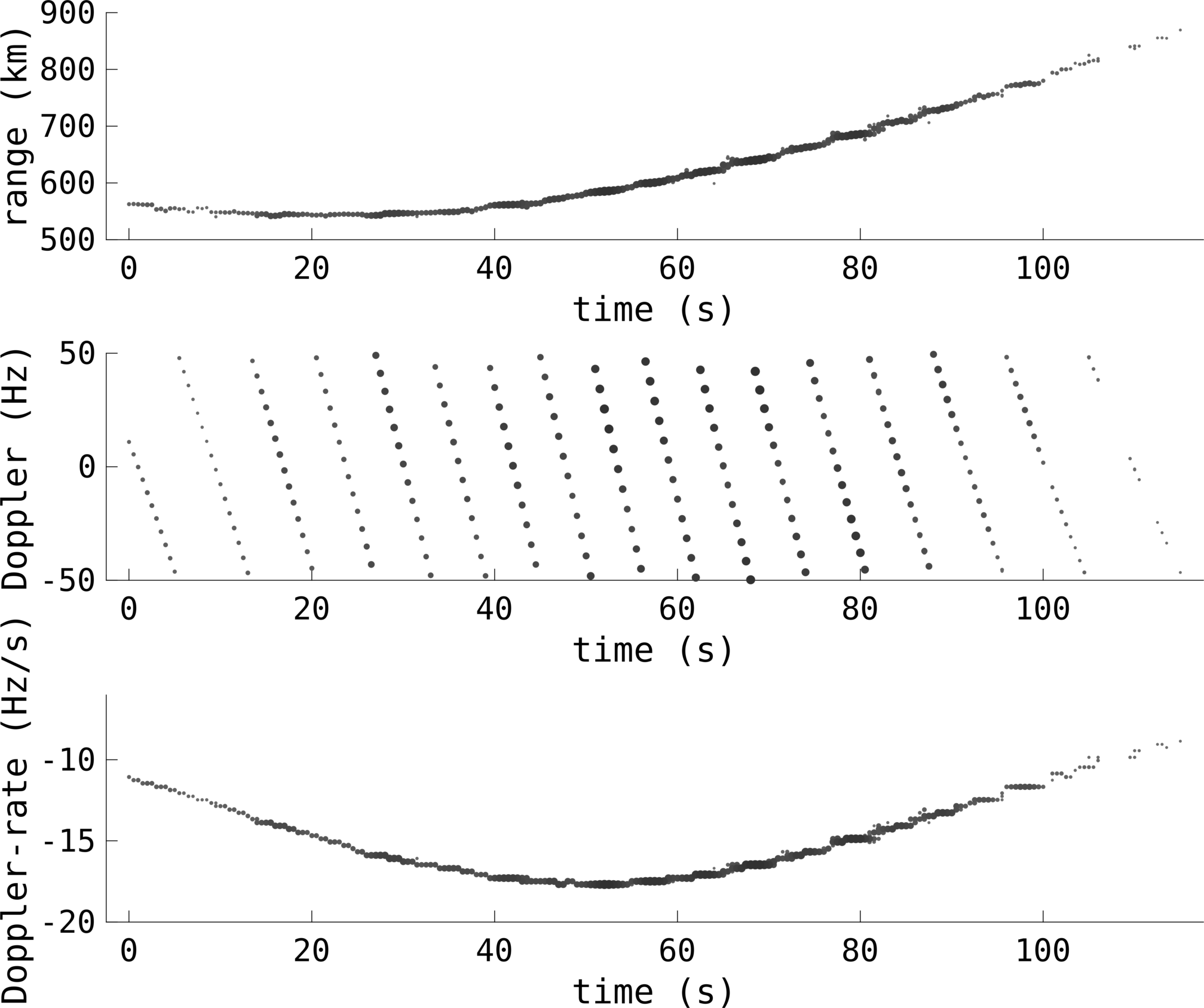}
\end{center}
\caption{Example range, Doppler, and Doppler-rate detections from a typical satellite pass detected by the HFLOS radar. The Doppler measurements are ambiguous and the range--Doppler coupling bias has not yet been removed. }
\label{fig:full_coupled_range_doppler_coupling}
\end{figure}

 Figure \ref{fig:full_coupled_range_doppler_coupling} shows the detections from a typical RSO pass, showing range, Doppler, and Doppler rate. As these detections occurred when the HFLOS radar was in a quasi-monostatic configuration, the true range is given, that is, half of the $r_m$ term in \eqref{eq:measured_bistatic_range}. The WRF was 100~Hz and the satellite was wrapped in Doppler 16 times. The CIT was 5.12~s and the inter-CIT stride time was 0.5~s. Having such a short stride time, along with accurate Doppler rate measurements, greatly assists in the unwrapping of the Doppler, particularly when  measurements are intermittent. The impact of the range--Doppler coupling is apparent in the lopsided nature of the range profile. The point of closest approach, the minimum range, when the Doppler rate is at its greatest magnitude, also corresponds to when the fold factor is zero and the Doppler is unambiguous.

\begin{figure}[ht]
\begin{center}
\includegraphics[width=\columnwidth]{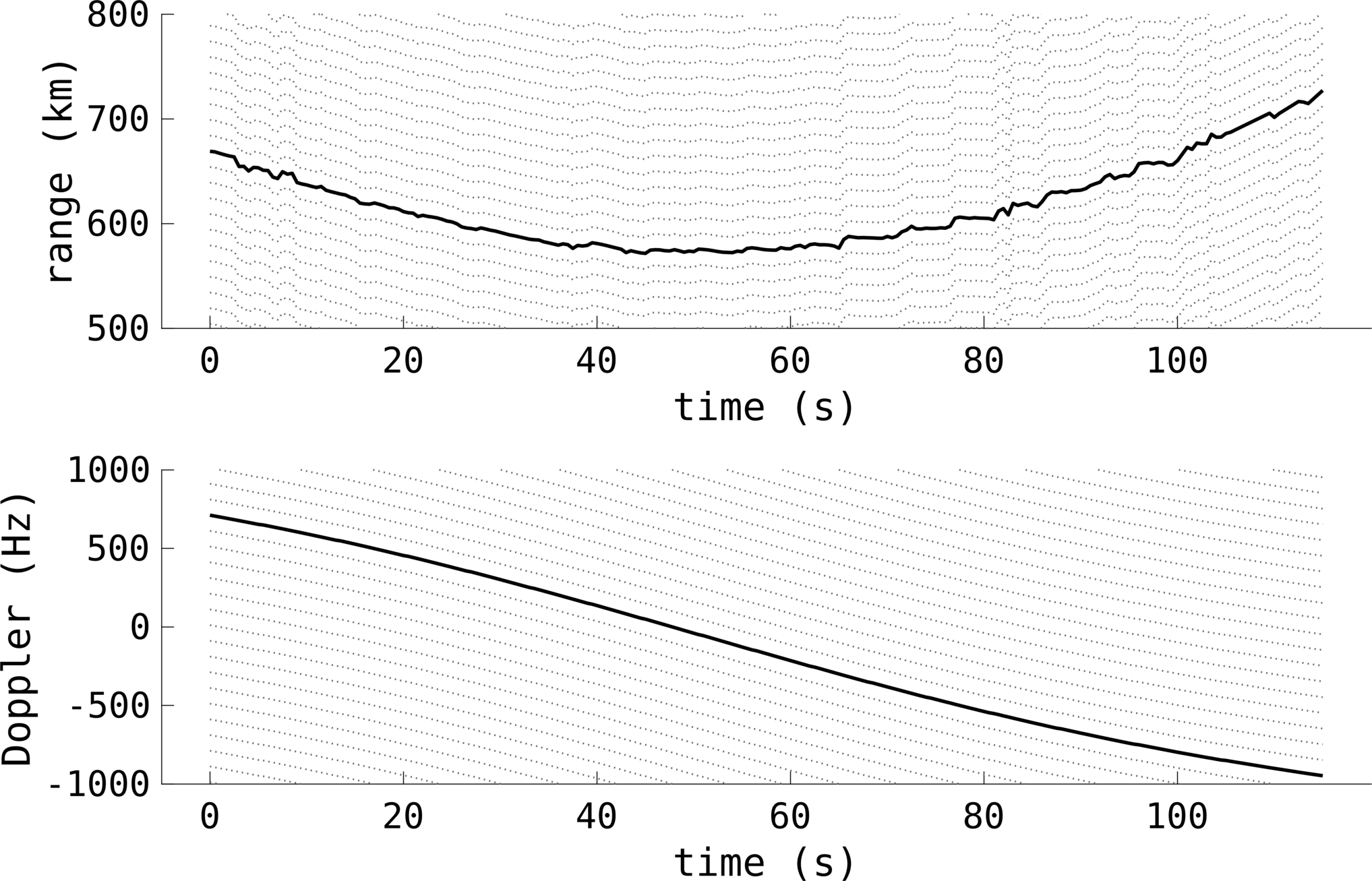}
\end{center}
\caption{The corrected range and Doppler measurements from Figure \ref{fig:full_coupled_range_doppler_coupling}. As well as the correct values (solid black), the potential ambiguous Doppler is also shown along with the corresponding range measurements (dotted).}
\label{fig:range_doppler_coupling}
\end{figure} 

Figure \ref{fig:range_doppler_coupling} contains the same detections as Figure \ref{fig:full_coupled_range_doppler_coupling}. However, now the Doppler has been correctly unwrapped and the range--Doppler coupling bias has been removed. The range--Doppler coupling correction has altered the ranges by almost 100~km at the beginning and end of the pass, correcting the previously lopsided range profile. Also shown in the figure is the large number of ambiguities, being only 100 Hz apart, highlighting the very narrow ambiguous window where these detections are formed. 

The detections also have an associated spatial pointing direction from the beamforming. For terrestrial purposes, the radar beamforms are in azimuth and elevation; however, these are not ideal for RSO tracking. Instead, direction cosines are used, as the RSO trajectory will be far simpler, more akin to straight lines, instead of the complicated curves that result from using azimuth and elevation.

\begin{figure}[ht]
\begin{center}
\includegraphics[width=\columnwidth]{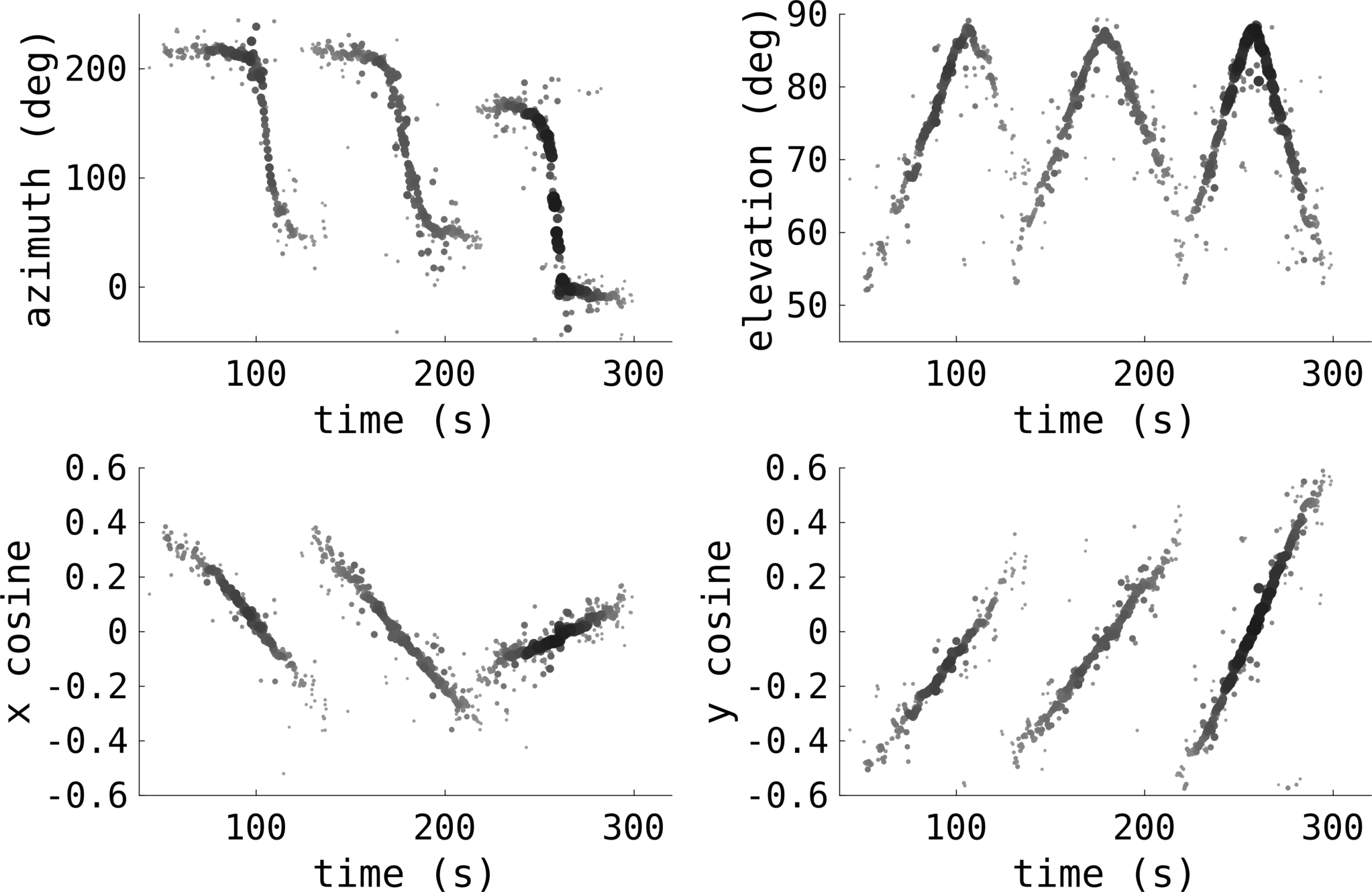}
\end{center}
\caption{Example spatial radar detections from the pass of three RSOs are shown, both azimuth and elevation as well as the direction cosines.}
\label{fig:super_sky_angles}
\end{figure}

The direction cosines used are relative to the zenith, $x$ for east--west and $y$ for south--north. If the azimuth is given by $\theta$ and the elevation is given by $\phi$, then the direction cosines are given by
\begin{IEEEeqnarray}{rCL}
x = && \sin{\theta}\cos{\phi}~, \label{eq:cosine_x}\\
y =& & \cos{\theta}\cos{\phi} ~.\label{eq:cosine_y}
\end{IEEEeqnarray} An example illustrating the benefit of these parameters is given in Figure \ref{fig:super_sky_angles}, showing the pass of three RSOs near the zenith, with both azimuth and elevation as well as the direction cosines. As the RSOs approach the azimuth discontinuity at the zenith, there is a sharp jump in both azimuth and elevation. In comparison, the cosines are relatively simple trajectories. This greatly assists in tracking, allowing simpler spatial motion models and reduced process noise.

Linear arrays are often used with HF radar, especially for low-elevation transmit illumination directed towards the horizon. When used in this manner, the spatial angle that the linear array observes is often treated as simply being equivalent to the azimuth. However, the single spatial dimension in which the linear array is able to discriminate is more complicated; it is a conflation of the azimuth and elevation, and is referred to as the `coning angle'~\cite{frazer2013orbit}. If the direction of the array axis is given by $\boldsymbol{\mu}_p$, parallel (or endfire) to the receive array in the east, north, up (ENU) Cartesian reference frame, then the coning angle, $\alpha$, is given by
\begin{IEEEeqnarray}{rCL}
\alpha &=& \arccos{\left( \boldsymbol{\mu}_p \cdot {\begin{bmatrix} \sin{\theta}\cos{\phi}, & \cos{\theta}\cos{\phi}, & \sin{\phi} \\ \end{bmatrix}}^T  \right)}~.~~~~~~
\end{IEEEeqnarray} This highlights the azimuth and elevation conflation, but the coning angle can also be expressed in terms of $\rhovec_{r}$, so long as the array axis is transformed to the appropriate reference frame.

The coning angle makes RSO localisation challenging, because unlike a two-dimensional array, a detection is not providing a location estimate. A measurement consisting of a range and coning angle defines a circle of potential locations for a monostatic radar, or an ellipse for bistatic radar~\cite{martin2019new}.

\subsection{Tracking}

As described in earlier work~\cite{frazer2013orbit}, for these HFLOS activities a Probabilistic Multi-Hypothesis Tracker (PMHT) algorithm is used for estimating target trajectories~\cite{streit1994maximum}. The PMHT algorithm is a method for finding the best estimate of the target states when the measurement source is unknown. Using an expectation maximisation algorithm, PMHT models the assignment of measurements to targets as hidden variables and estimates target states by taking the expectation over these assignments. Unlike conventional data association techniques, PMHT has the advantage of linear complexity in terms of the number of detections per CIT, the number of CITs per batch, and the number of targets by assuming independence between association hypotheses.

Tracks are initiated using a two-detection initiation scheme, where the PMHT algorithm predicts the future target states, associates measurements with tracks, and updates track states based on the weighted chosen measurements. An overarching layer is also required to monitor and manage the tracks, controlling the number of tracks through culling and promotion, etc., to ensure quality output. The ultimate result, after processing and filtering, is a series of tracks alongside associated detections, to be used for space surveillance.

\section{Correction for Ionospheric Effects}\label{sec:sec4}
Interaction between HF radio waves and free electrons in the ionosphere results in ionospheric refraction. The effects of this on an HFLOS RSO observation are threefold: retardation of the group velocity alters the measured range, advance of the phase velocity alters the measured Doppler, and bending of the radio waves alters the measured angular position~\cite{guier1959doppler,10371125}.

An algorithm has been developed that performs three-dimensional {magneto-ionic} numerical ray tracing, using the ray tracing package PHaRLAP~\cite{cervera2014modeling}, through the International Reference Ionosphere (IRI) model 2020~\cite{bilitza2022international,7775262}, to calculate the ionospheric
correction to an HFLOS RSO observation~\cite{hflos_iono_corrections_tristan}.

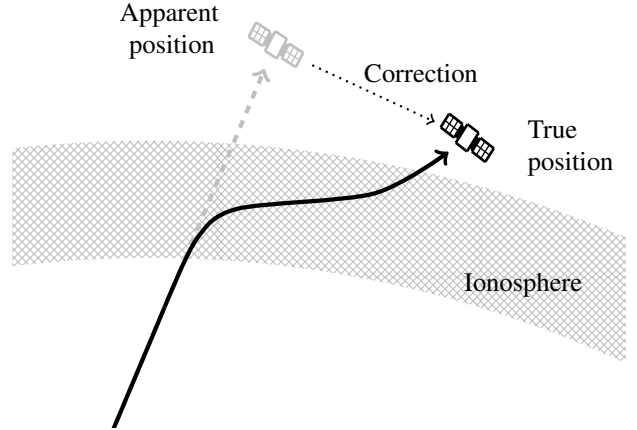
\begin{figure}[!ht]
\begin{tikzpicture}[dot/.style={draw,fill,circle,inner sep=1pt},xscale=0.9]
\clip (0,-0.5) rectangle + (9,5.75); 

    \draw[pattern color=lightgray, pattern=crosshatch, draw=none] (2,-16.5) circle [radius=19.8];
    \draw[fill=white, draw=none] (2,-16.5) circle [radius=18.25];

    \draw [-To,lightgray,ultra thick,dashed] (1.25,-1 ) -- (3.7,4.25) ;

    \draw [black,ultra thick] (1.25,-1 ) -- (2.23,1.1) ;

    \draw [-To,black,ultra thick] plot [smooth] coordinates { (2.23,1.1)  (2.7, 2)  (3.3,2.4) (5.3,2.6) (6.4,3.15)};

    \begin{scope}[xshift=190 , yshift=95, rotate=-35, scale=0.1]
    \draw[line width=1pt, color=black](-1, 1.5) -- (1, 1.5); 
    \draw[line width=1pt, color=black](1, 1.5) -- (1, -1.5);
    \draw[line width=1pt, color=black](1, -1.5) -- (-1, -1.5);
    \draw[line width=1pt, color=black](-1, -1.5) -- (-1, 1.5);

    \draw[line width=1pt, color=black](-4, 1.1) -- (-1.6, 1.1); 
    \draw[line width=1pt, color=black](-1.6, 1.1) -- (-1.6, -1.1);
    \draw[line width=1pt, color=black](-1.6, -1.1) -- (-4, -1.1);
    \draw[line width=1pt, color=black](-4, -1.1) -- (-4, 1.1);

    \draw[line width=3pt, color=black](-1.6, 0.4) -- (-1, 0.4);
    \draw[line width=3pt, color=black](-1.6, -0.4) -- (-1, -0.4);

    \draw[line width=3pt, color=black](1.6, 0.4) -- (1, 0.4);
    \draw[line width=3pt, color=black](1.6, -0.4) -- (1, -0.4);

    \draw[line width=1pt, color=black](4, 1.1) -- (1.6, 1.1);
    \draw[line width=1pt, color=black](1.6, 1.1) -- (1.6, -1.1);
    \draw[line width=1pt, color=black](1.6, -1.1) -- (4, -1.1);
    \draw[line width=1pt, color=black](4, -1.1) -- (4, 1.1);

    \draw[line width=0.5pt, color=black](1.6, 0) -- (4, 0);
    \draw[line width=0.5pt, color=black](2.4, -1.1) -- (2.4, 1.1);
    \draw[line width=0.5pt, color=black](3.2, -1.1) -- (3.2, 1.1);

    \draw[line width=0.5pt, color=black](-1.6, 0) -- (-4, 0);
    \draw[line width=0.5pt, color=black](-2.4, -1.1) -- (-2.4, 1.1);
    \draw[line width=0.5pt, color=black](-3.2, -1.1) -- (-3.2, 1.1);

    \end{scope}

    \begin{scope}[xshift=110 , yshift=130, rotate=-30, scale=0.1]
    \draw[line width=1pt, color=lightgray](-1, 1.5) -- (1, 1.5); 
    \draw[line width=1pt, color=lightgray](1, 1.5) -- (1, -1.5);
    \draw[line width=1pt, color=lightgray](1, -1.5) -- (-1, -1.5);
    \draw[line width=1pt, color=lightgray](-1, -1.5) -- (-1, 1.5);

    \draw[line width=1pt, color=lightgray](-4, 1.1) -- (-1.6, 1.1); 
    \draw[line width=1pt, color=lightgray](-1.6, 1.1) -- (-1.6, -1.1);
    \draw[line width=1pt, color=lightgray](-1.6, -1.1) -- (-4, -1.1);
    \draw[line width=1pt, color=lightgray](-4, -1.1) -- (-4, 1.1);

    \draw[line width=3pt, color=lightgray](-1.6, 0.4) -- (-1, 0.4);
    \draw[line width=3pt, color=lightgray](-1.6, -0.4) -- (-1, -0.4);

    \draw[line width=3pt, color=lightgray](1.6, 0.4) -- (1, 0.4);
    \draw[line width=3pt, color=lightgray](1.6, -0.4) -- (1, -0.4);

    \draw[line width=1pt, color=lightgray](4, 1.1) -- (1.6, 1.1);
    \draw[line width=1pt, color=lightgray](1.6, 1.1) -- (1.6, -1.1);
    \draw[line width=1pt, color=lightgray](1.6, -1.1) -- (4, -1.1);
    \draw[line width=1pt, color=lightgray](4, -1.1) -- (4, 1.1);

    \draw[line width=0.5pt, color=lightgray](1.6, 0) -- (4, 0);
    \draw[line width=0.5pt, color=lightgray](2.4, -1.1) -- (2.4, 1.1);
    \draw[line width=0.5pt, color=lightgray](3.2, -1.1) -- (3.2, 1.1);

    \draw[line width=0.5pt, color=lightgray](-1.6, 0) -- (-4, 0);
    \draw[line width=0.5pt, color=lightgray](-2.4, -1.1) -- (-2.4, 1.1);
    \draw[line width=0.5pt, color=lightgray](-3.2, -1.1) -- (-3.2, 1.1);
    \end{scope}

    \node at (7.5,1.4) {Ionosphere};

    \node[align=left] at (8.2,3.2) {True\\position};

    \node[align=left] at (2.3,4.75) {Apparent\\~position};

    \draw [-To,black,thick,dotted] (4.4,4.3 ) -- (6.2,3.55) ;

    \node at (6,4.2) {Correction};

\end{tikzpicture}

\caption{Illustration of the ray-bending that occurs, resulting in the apparent position of an RSO differing from the true position, requiring a correction. At the upper end of the HF band, the amount of ray-bending is small. It is exaggerated here for illustrative
purposes.} \label{fig:ionosphere_shift}
\end{figure}

The position of the RSO is corrected first, attempting to overcome the ray-bending illustrated in
Figure \ref{fig:ionosphere_shift}. At each step along the RSO’s path, the measured elevation and
azimuth angles are used as the starting angles for a ray modelled by PHaRLAP.
This ray is traced through the model ionosphere until the modelled
group range matches the measured range, at which point the ray tracing
is terminated. The modelled ray accounts for group retardation and ray-bending, and the location of the end point of the ray is the modelled true
position of the RSO, assuming the model ionosphere is accurate. This is then
converted to a corrected range, range rate, elevation, and azimuth.

Two different methods are used to calculate the correction to the range
rate (Doppler). The first, referred to as the Raytracing method, calculates the
corrected range rate by taking the numerical derivative of the corrected range, calculated above: 
\begin{IEEEeqnarray}{rCL}
RR_c = \frac{dR_c}{dt},
\label{corrected}
\end{IEEEeqnarray}
where $RR_c$ is the corrected range rate and $R_c$ is the raytraced corrected range.

However, random fluctuations in the radar range measurements are replicated
in the corrected range. Calculating the numerical derivative can therefore
lead to uncertainties becoming unsuitably large. A second method, referred
to as the Phase-Path Correction method was developed to mitigate this.
Consider the measured Doppler shift, $\delta f$~\cite{bennett1967calculation}:
\begin{IEEEeqnarray}{rCL}
\delta f = -2\frac{f}{c}\frac{dP}{dt},
\end{IEEEeqnarray}
where $f$ is the radio wave frequency, $c$ the speed of light in vacuo, $P$ is the phase path, and the factor of 2 accounts for propagation both from the radar to the RSO and back.
As range rate is related to Doppler shift by $-\frac{c\delta f}{2f}$, we have the following equation for the measured range rate:
\begin{IEEEeqnarray}{rCL}
\label{measured}
RR_m= \frac{dP}{dt}.
\end{IEEEeqnarray}
Since the phase path is also calculated by PHaRLAP during the ray tracing process,
the phase path correction, $\delta P = R_c - P$ can be calculated independently of the range rate measurement.
In this expression, taking the difference between $R_c$ and $P$ effectively subtracts any errors which were replicated from the measured range.
We can now take the derivative:
\begin{IEEEeqnarray}{rCL}
\frac{d(\delta P)}{dt} = \frac{dR_c}{dt} - \frac{dP}{dt}.
\end{IEEEeqnarray}
Substituting in the expressions in Equations \ref{corrected} and \ref{measured}, we have a second, independent method of calculating the corrected range rate:
\begin{IEEEeqnarray}{rCL}
RR_c = RR_m + \frac{d(\delta P)}{dt}.
\end{IEEEeqnarray}



\begin{figure}[ht]
\begin{center}
\includegraphics[width=\columnwidth]{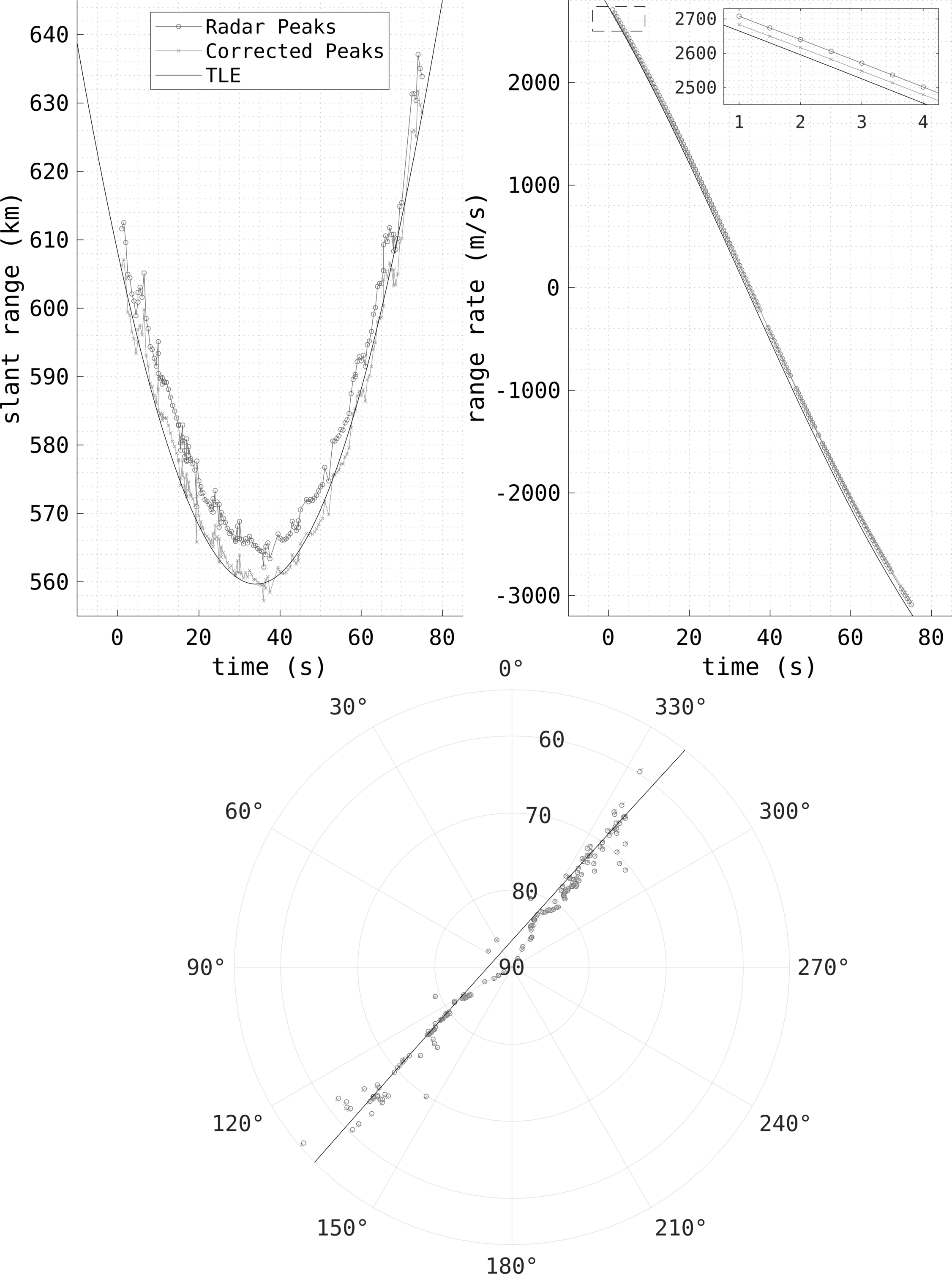}
\end{center}
\caption{HFLOS  range, {range rate (Doppler)}, and spatial observations of the pass of a Starlink satellite, before and after ionospheric corrections, with the associated TLE plotted for comparison.}
\label{fig:Starlink-1202}
\end{figure} 

Figure \ref{fig:Starlink-1202} shows the detections of the HFLOS observation of a Starlink satellite, before and after ionospheric correction. The range error has been significantly reduced, with the detections now better matching the two-line element (TLE). There are also slight corrections to the Doppler and spatial parameters to improve the measurements. As the data were collected near solar minimum, the impact of the ionosphere on the radar measurements is considerably less than during solar maximum~\cite{henault2025space}. {These methods are able to significantly improve the range measurements, with the mean range measurement error being reduced from 5~km to 1~km~}\cite{hflos_iono_corrections_tristan}{, albeit with a large standard deviation due to the coarse range resolution. The other measurement parameters are not significantly improved; however,} even slight improvements to the measurement accuracy may greatly improve the resulting orbital accuracy.

\section{Space Surveillance}\label{sec:sec5}

\begin{figure*}[ht]
\begin{center}
\includegraphics[width=\textwidth]{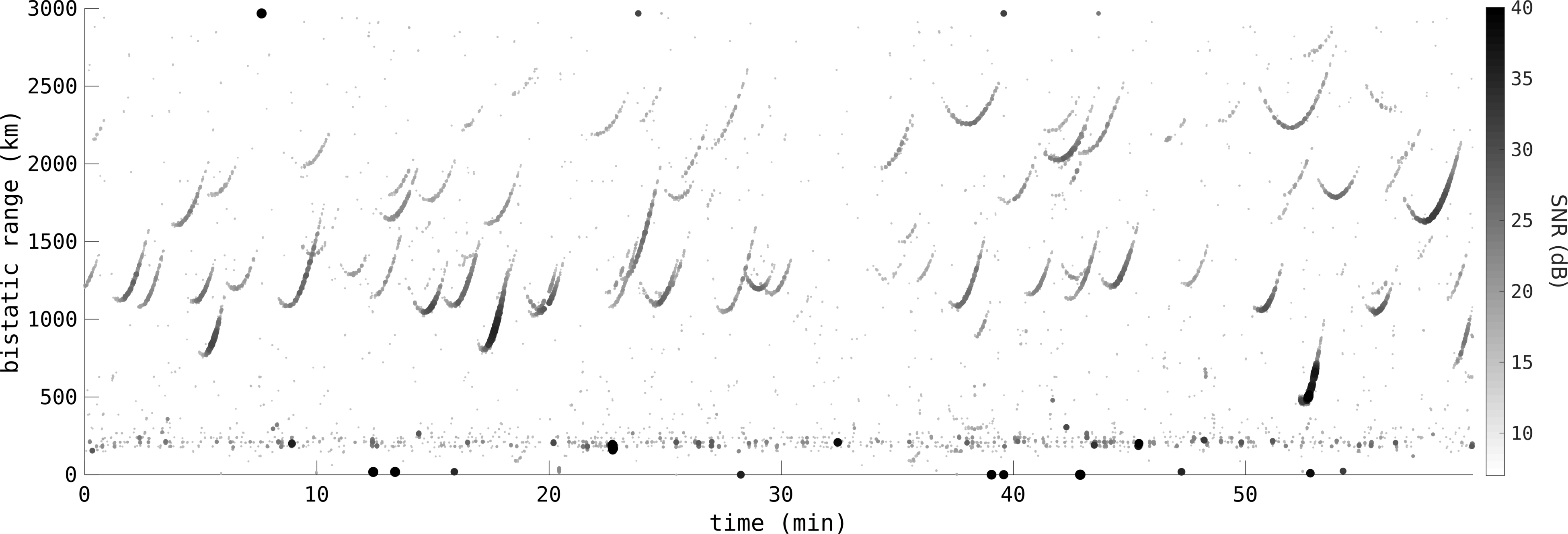}
\end{center}
\caption{Range vs time profile of one hour of detections observed by the HFLOS radar. Only detections associated with Doppler-rate processing are shown. Many satellites are visible as well as a meteor band between 100 and 250 km. {These observations occurred in 2018 and predate much of the recent rise in the number of RSOs in LEO.}}
\label{fig:one_hour_detections}
\end{figure*} 

SpaceFest~\cite{SpaceFestArticle} is a series of Australian space surveillance campaigns that have been run, in some form, since 2018.~
The HFLOS system  participated in SpaceFest on numerous occasions between 2018 and 2022,  deploying to South Australia's Far North and conducting space surveillance activities in conjunction with other sensors. These repeated deployments allowed for iterative system development and verification, conducting experiments with new hardware, array configurations, and signal processing. Receive array deployments included planar arrays (as shown in Figure \ref{fig:hflos_rx}), sparse circular arrays, multiple linear arrays, and general, large-baseline, multistatic configurations.

The HFLOS radar was able to perform real-time wide-area surveillance, producing accurate orbital estimates, including the cueing of narrow field-of-view sensors. \mbox{Figure \ref{fig:one_hour_detections}} shows the range vs time profile of one hour of RSO passes observed by HFLOS, {with a 90\textdegree~transmitter beamwidth illumination, a bandwidth of 10~kHz, and a PRF of 100~Hz}. Only the detections formed from Doppler-rate processing are shown. Despite the Doppler-rate filtering, there are still a number of meteor detections apparent in a band at around 200~km range. Again, the impact of range--Doppler coupling is apparent in the range profiles. Instead of a typical parabola-like shape, the range profile of each RSO pass appears more like a tick. 
A satellite initially has significant negative radial velocity as it appears above the horizon, to a point of closest approach where the radial velocity is (instantaneously) zero, through to a point where the radial velocity is very positive. This large radial velocity will bias the range measurements in a direction depending on the polarity of the chirp (per \eqref{eq:measured_bistatic_range}).  It is apparent that an upchirp LFM was used because the approaching ranges are reduced with the negative velocity whilst the receding ranges are increased with the positive velocity. Only a relatively small number of noise and other detections make it through the Doppler-rate filtering, making tracking far easier.

\begin{figure*}[ht]
\begin{center}
\setlength{\fboxrule}{0pt}  
    \fbox{\includegraphics[width=.3175\textwidth]{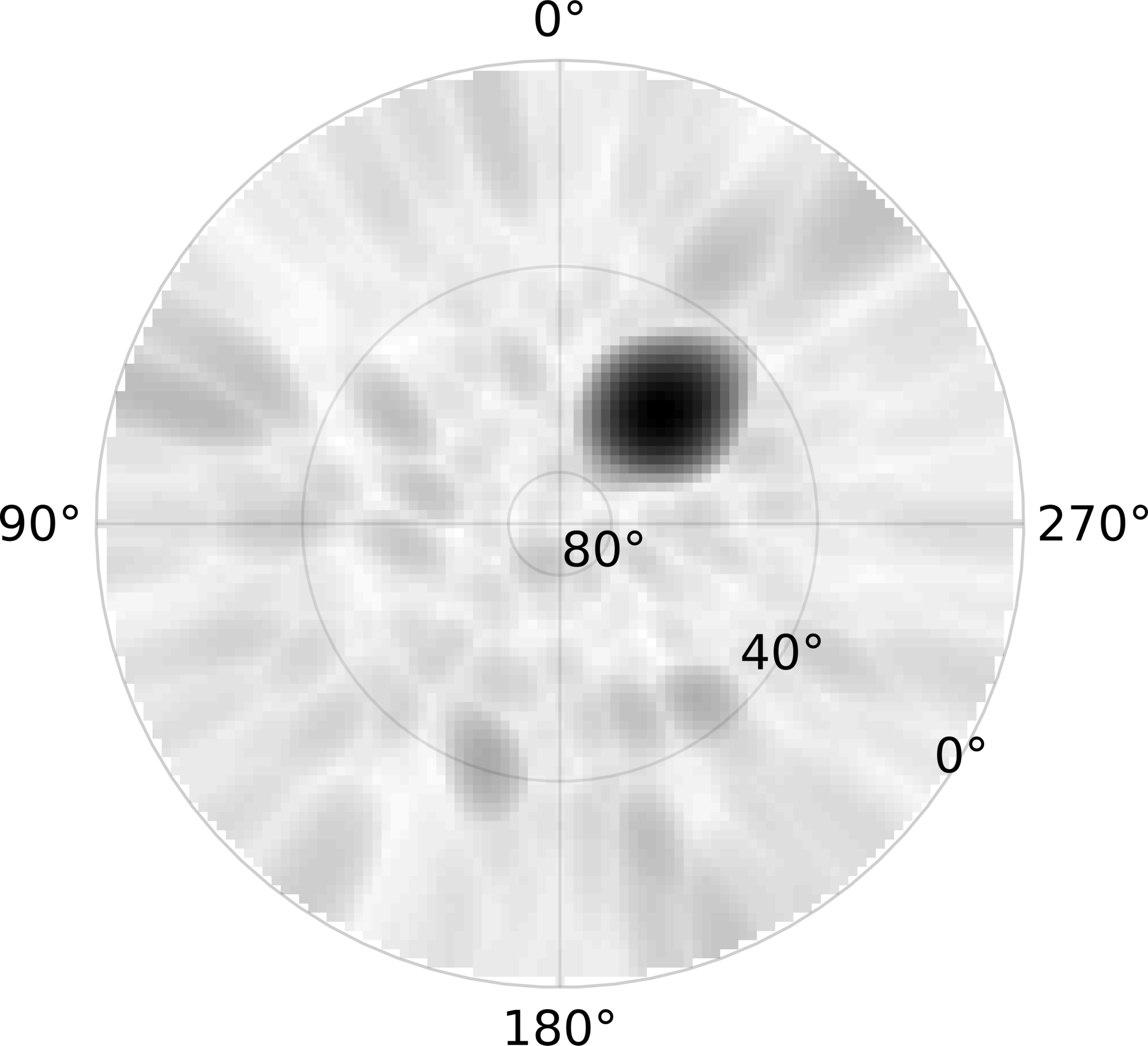}}   
    \fbox{\includegraphics[width=.3175\textwidth]{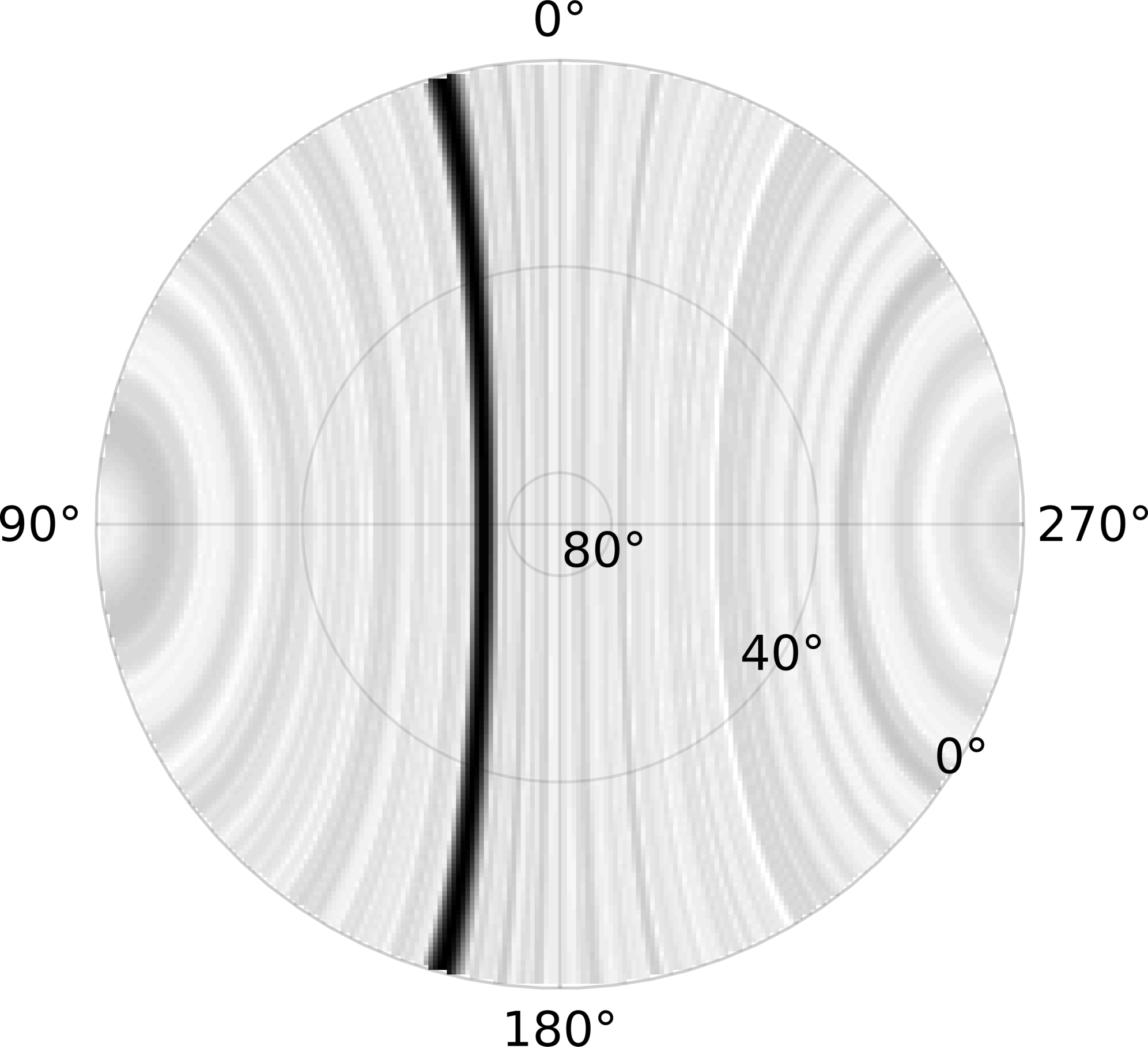}} 
    \fbox{\includegraphics[width=.3175\textwidth]{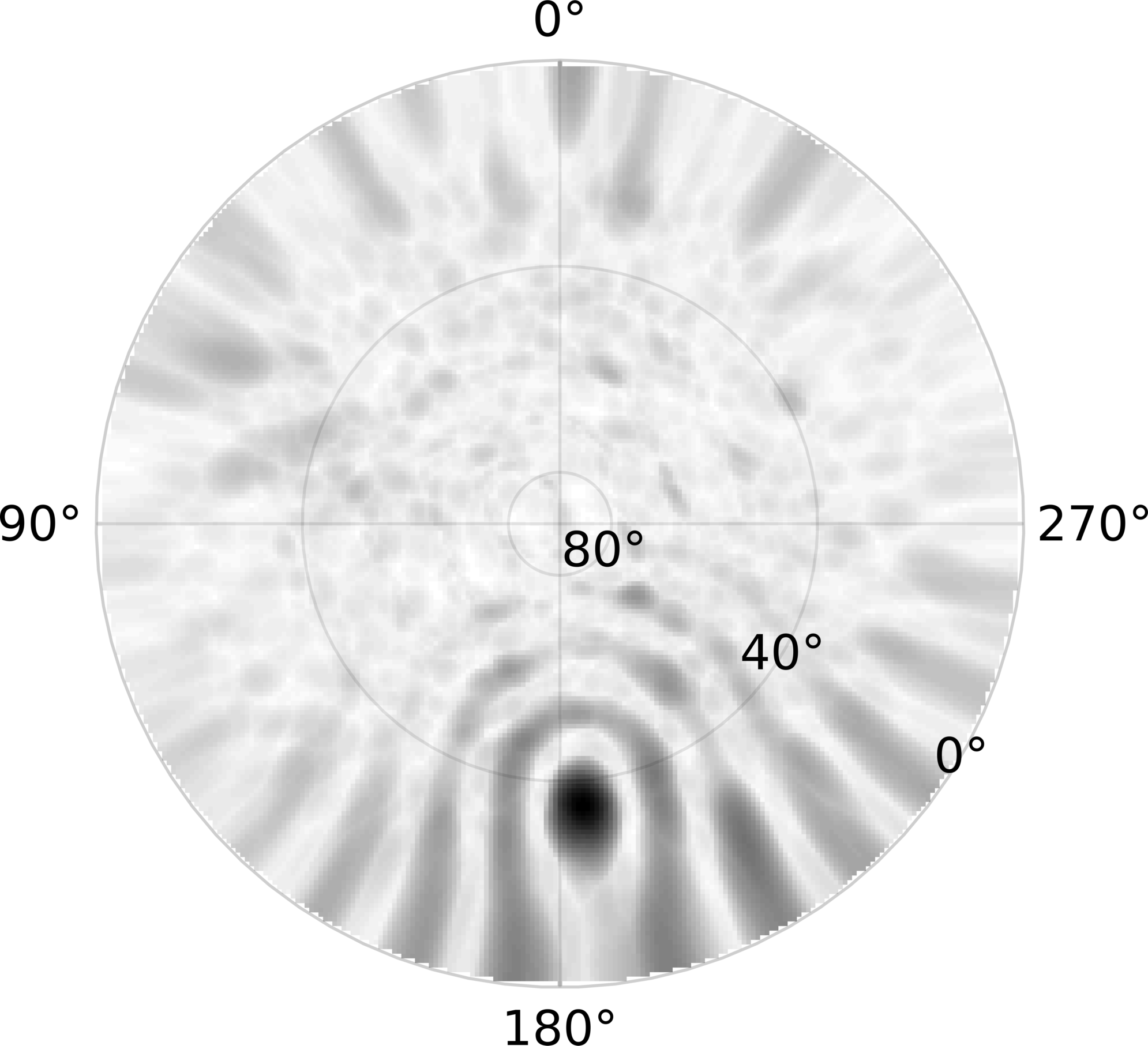}} 
\end{center}
\caption{The two-dimensional spatial returns for three RSOs observed by three different receive array configurations. The left subplot is from a compact planar array, the middle subplot from a linear array, and the right from  sparse circular array.}
\label{fig:array_factor}
\end{figure*}

Many receive array configurations have been utilised in these trials to meet the specific goals of each deployment. Figure \ref{fig:array_factor} shows the spatial pattern of an RSO, highlighting the array factor, for three receiver configurations: a compact planar array, a linear array, and a sparse circular array, respectively. {The planar array was a two-dimensional staggered 5x6 element grid with each antenna separated from adjacent elements by 5~m, the linear array consisted of 30 elements with a 5.2~m element spacing, and the circular array consisted of 32 elements equally spaced on forming a 50~m diameter circle. These configurations were chosen to give an element spacing of approximately half a wavelength at the frequencies of interest.} {The three patterns in Figure }\ref{fig:array_factor}{,} indicative of the array factor, highlight the trade-space between measurement resolution, sidelobe levels, and structure, as well as highlighting the localisation challenges of using a linear array. {With the element spacing being approximately half a wavelength, the planar array produces unambiguous spatial estimates; however, the large beamwidth results in coarse detection localisation. Conversely, the circular array being much larger, and essentially sparse (as the circle is not filled), produces higher accuracy spatial estimates, but also produces larger, and structured, sidelobes. These sidelobes are often not an issue, as the main peak is easily found; however, they can present issues when RSOs occupy similar delay and Doppler bins but in different directions (which is not uncommon in orbital regimes occupied by a large number of satellites, such as LEO communication mega constellations).}   The linear array can only discriminate in one spatial direction, and the RSO pattern spans a large range of azimuths, from 15\textdegree~through to 165\textdegree,~and elevation from the horizon up to 70\textdegree. Note that the one-dimensional linear array would not ordinarily be processed as a two-dimensional sky map; this is simply demonstrating the coning angle.

Other experiments that have been conducted have included the real-time multistatic processing of radar returns, including accurate geolocation from two co-located perpendicular linear arrays, the use of two transmitted signals with opposite chirp polarities for instantaneous accurate fold-factor estimation, and the networked dissemination of radar data, detections, and tracks for broad fusion with other sensors.

The signal processing stage provides tracks, as well as each track's associated detections. The tracking process simply associates, and maintains custody of, the detections, and does not necessarily provide a precise or accurate description of the orbital position and velocity along the track. Instead, the track's associated detections are used in a subsequent OD stage to best estimate the true orbital trajectory of the RSO. This is the broad aim of any space surveillance sensor---to determine the orbit accurately enough so that the track can be observed (and associated) again, either by the same sensor or others in a network~\cite{Vallado2001fundamentals}.

\begin{figure*}[ht]
\begin{center}
\scalebox{0.9}{
\tikzstyle{block} = [rectangle, draw, text centered, rounded      corners, minimum height=3em]
\begin{tikzpicture}[xshift=-10]
\begin{scope}[xshift=-10]
\node (n1) at (-12,0) [block]  {Detections};
\node (n2) at (-10,0) [block] {Track};
\node (n3) at (-8,0) [block] {TLE match?};
\node (n4) at (-4.25,0) [block] {Existing track match?};
\node (n5) at (-1.2,0) [block] {IOD};
\node (n6) at (0,0) [block] {OD};
\node (n7) at (0,-2) [block] {Update track store};
\draw [->] (n1.east) -- (n2.west);
\draw [->] (n2.east) -- (n3.west);
\draw [->] (n3.east) -- (n4.west);
\draw [->] (n4.east) -- (n5.west);
\draw [->] (n5.east) -- (n6.west);
\draw [->] (n3.south) -| ++(0,-1) |- (n7.west);
\draw [->] (n4.south) -| ++(0,-1) |- (n7.west);
\draw [->] (n6.south) -- (n7);
\node () at (-6.7, 0.275) {\small{N}};
\node () at (-2.2, 0.275) {\small{N}};
\node () at (-7.7, -.8) {\small{Y}};
\node () at (-4, -.8) {\small{Y}};
\end{scope}
\end{tikzpicture}
}
\end{center}
\caption{{Diagram showing the track management process; tracks are compared against the known TLE catalogue as well as existing tracks before being used to generate initial orbits and new orbital estimates.}\label{fig:flow_chart}}
\end{figure*}
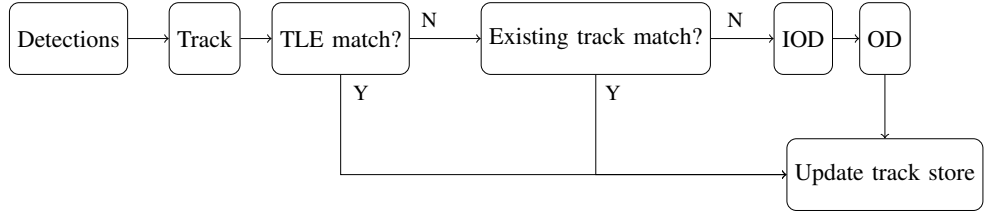

{Figure }\ref{fig:flow_chart}{ outlines the broad track management approach and indicates when the IOD and OD steps are required. A given track's associated detections are used to determine if the detections are from a known object in the space catalogue; if not, they are used to determine if they match an object detected previously, and failing that, the detections are used to form an orbital estimate that can then be used to compare against subsequent passes.  }

{The initial step is to associate the track with any already-known RSOs, provided by catalogues such as the Space-Track} service maintained by the Space Surveillance Network~\cite{USSPACECOM}. This service maintains a comprehensive catalogue of objects in Earth orbit, and these tracks are published as TLE sets. The TLE is a text-based description of an Earth-centered orbital ellipse, as well as the location of an RSO on the ellipse. The six parameters defining the ellipse are a sufficient description of a two-body orbit (equivalent to the six-Cartesian-element position and velocity description). The TLE, however, contains extra information about the orbit, such as the parameters for a simple drag model, so that the orbit can be more accurately propagated using the Simplified General Perturbations model, Version 4. The TLEs are used to verify how precise the HFLOS-predicted orbits are, although it is worth {noting} that the TLEs are only maintained to a limited level of accuracy themselves.

TLEs are associated with the track's measurements by comparing the expected measurements from the TLE against the track's coupled bistatic range, bistatic Doppler rate, and the spatial parameters. Wrapped Doppler is typically not used, as the differences between the TLE and the measurements are not very meaningful until the Doppler is unwrapped. Associating a track prior to range--Doppler coupling correction and Doppler unwrapping allows for very short tracks to be associated, without the risk of unwrapping or correction errors resulting in a missed association.

\begin{figure}[ht]
\begin{center}
\includegraphics[width=\columnwidth]{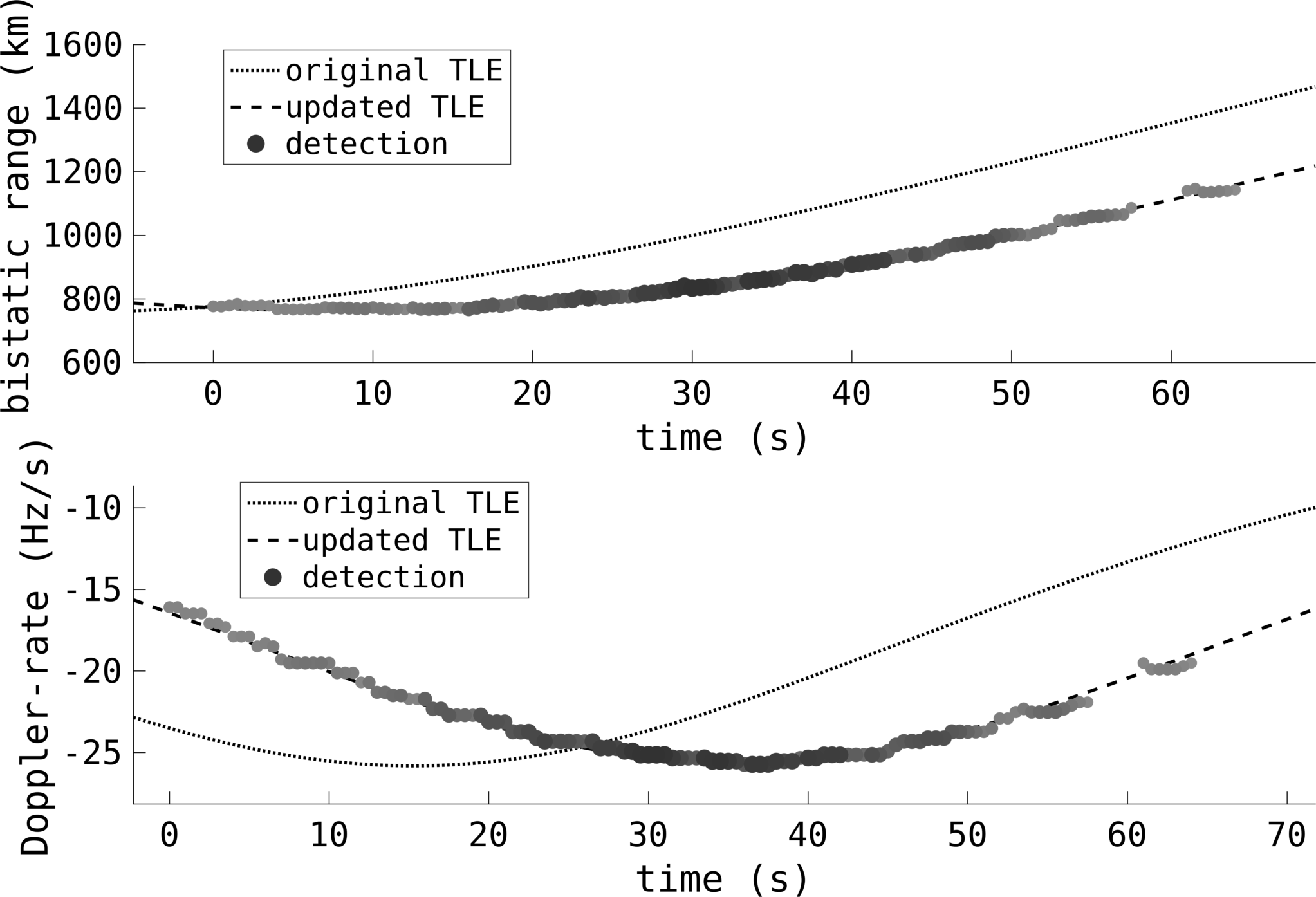}
\end{center}
\caption{The HFLOS range and Doppler-rate measurements of a Starlink satellite are shown, as well as the associated TLE before and after the TOES adjustment. The detections of the satellite are consistent with a 20~s TOES adjustment.}
\label{fig:SUPER_tle_TOES_shift_20s_norad_46551}
\end{figure}

One of the main benefits of wide-area surveillance is the detection and association of manoeuvring objects, or when the prior track information is outdated and stale. Narrow field-of-regard sensors, or sensors requiring a priori track information, can miss these targets. Figure \ref{fig:SUPER_tle_TOES_shift_20s_norad_46551} shows the association of a Starlink track with its TLE, despite the deviation. The automatic association is achieved by not only searching for a TLE match against the measurements, but time off element set (TOES) search for each TLE as well~\cite{Vallado2001fundamentals}. The TOES for each TLE is adjusted, forward and back, to find the best match. Note that the TOES adjustment is not simply shifting the TLE-derived measurements forward or back in time; instead, the TLE's mean anomaly is updated by an appropriate amount, determined by the time offset and the mean motion. The shifted TLE will not be a good estimate of the track's true orbit; it is merely a method for the association of RSOs that might be undergoing in-track boosting or experiencing significant in-track drag. If the RSO is manoeuvring out of plane, then such TOES estimates may not be sufficient for associating tracks.

The track's Doppler is then unwrapped and the range--Doppler coupling bias is removed. Then, an IOD stage is used to determine a preliminary orbital estimate. The IOD stage is important for several reasons, which will be detailed below in Section \ref{ssec:iod}. However, the most important aspect of the IOD step is providing the initial value for the numerical estimation technique to perform the full OD estimate.

{Improved OD can be achieved if the detections from multiple passes can be used, as the span of the detections will cover hours or days instead of minutes. Multiple tracks can be associated by TLE association, or through a search of prior tracks~}\cite{pirovano2020data}{. For the experimental configurations described here, a constrained brute-force search is applied to compare two tracks by predicting the respective tracks forward and back, comparing the orbital elements, and attempting multi-pass OD to determine if they are associated. Whilst this approach may suffice for these short experimental campaigns, it would quickly become unfeasible for any persistent space surveillance system, and more advanced methods are needed for operational systems~}\cite{pastor2023track}{. Even if a track's orbital velocity estimate is not sufficiently accurate, many crucial parameters, such as orbital period, orbital speed, and eccentricity, can be used to drastically constrain, or prune, the large search space. However, this typically only holds for tracks resulting from a two-dimensional array, as velocity uncertainty is often greater with linear arrays, making track association more challenging.}

\section{Orbit Determination} \label{sec:sec6}
The OD stage determines the orbital position and velocity that, when propagated, minimises the residuals between the predicted orbit and the measurements. For a single pass, the use of a two-body model is sufficient. The model assumes that the only force acting on the RSO is the gravity due to Earth, and an orbit is able to be perfectly propagated from a single position and velocity. In the Earth-centered inertial (ECI) reference frame, if the radar receiver is at position {$\boldsymbol{r}_{r}$} 
, and it has rotational velocity and acceleration $\dot{\boldsymbol{r}}_{r}$ and $\ddot{\boldsymbol{r}}_{r}$, respectively, then slant-range vector and its time derivatives are given by  $\rhovec_{r} = \boldsymbol{r} - \boldsymbol{r}_{r}$, $\rhodvec_{r} = \dot{\boldsymbol{r}} - \dot{\boldsymbol{r}}_{r}$, and $\rhoddvec_{r} = \ddot{\boldsymbol{r}} - \ddot{\boldsymbol{r}}_{r}$ (and likewise for the transmitter's respective vectors). Here, $\boldsymbol{r}$ and $\dot{\boldsymbol{r}}$ are the orbital position and velocity (the sixtuple forming the description of the orbit), and the acceleration is due to Earth's gravity, dependant on the position and the standard gravitational parameter for Earth, $\mu$:

\begin{IEEEeqnarray}{rCL}
\boldsymbol{\ddot{r}}= -\frac{\mu}{\lvert\boldsymbol{r}\rvert^3}\boldsymbol{r}\label{eq:eci_acceleration},
\end{IEEEeqnarray}
With this simplest orbital model, all the measurement parameters can be calculated and estimated from a single three-dimensional position and velocity. That is, now Equations \eqref{eq:slant_range}--\eqref{eq:doppler_rate_hectic} and \eqref{eq:cosine_x}--\eqref{eq:cosine_y} solely depend on the initial orbit, along with the known location and motion of the sensor, as well as the radar parameters. This greatly simplifies the OD stage. {There are other more accurate orbital models (as discussed further in Section }\ref{ssec:multipass}{); however, these do not tend to improve the orbital estimate for a single pass, and the detection-level inaccuracy is far greater than the relatively smaller impact of the non-Keplerian orbits.}

If $\boldsymbol{z}$ is the set of all measurements for a track, $\boldsymbol{x}$ is the position and velocity sixtuple orbit description, and $\boldsymbol{f}$ is the measurement observation function, which maps an orbit to the measurement space at all times of interest, then the OD solution is given by:

\begin{IEEEeqnarray}{rCL}
\hat{\boldsymbol{x}} = \mathop{\arg\!\min}\limits_{\boldsymbol{x}}(\lvert\boldsymbol{z} - \boldsymbol{f}(\boldsymbol{x})\rvert^2)~. \label{eq:od_minimum}
\end{IEEEeqnarray}

As $\boldsymbol{f}$ is highly non-linear, finding a global minimum is typically not achievable, and instead a solution is found by linearising all quantities around an initial orbit estimate (the IOD estimate) and refining numerically~\cite{montenbruck2012satellite, hennessy2022establishing}. This orbital output is compared against the TLE, as the notional {truth}. The covariance of the OD process is also used to provide confidence in the generated orbit; this is generally useful, but it is especially so for when there is no associated TLE.

\begin{figure}[ht]
\begin{center}
\includegraphics[width=\columnwidth]{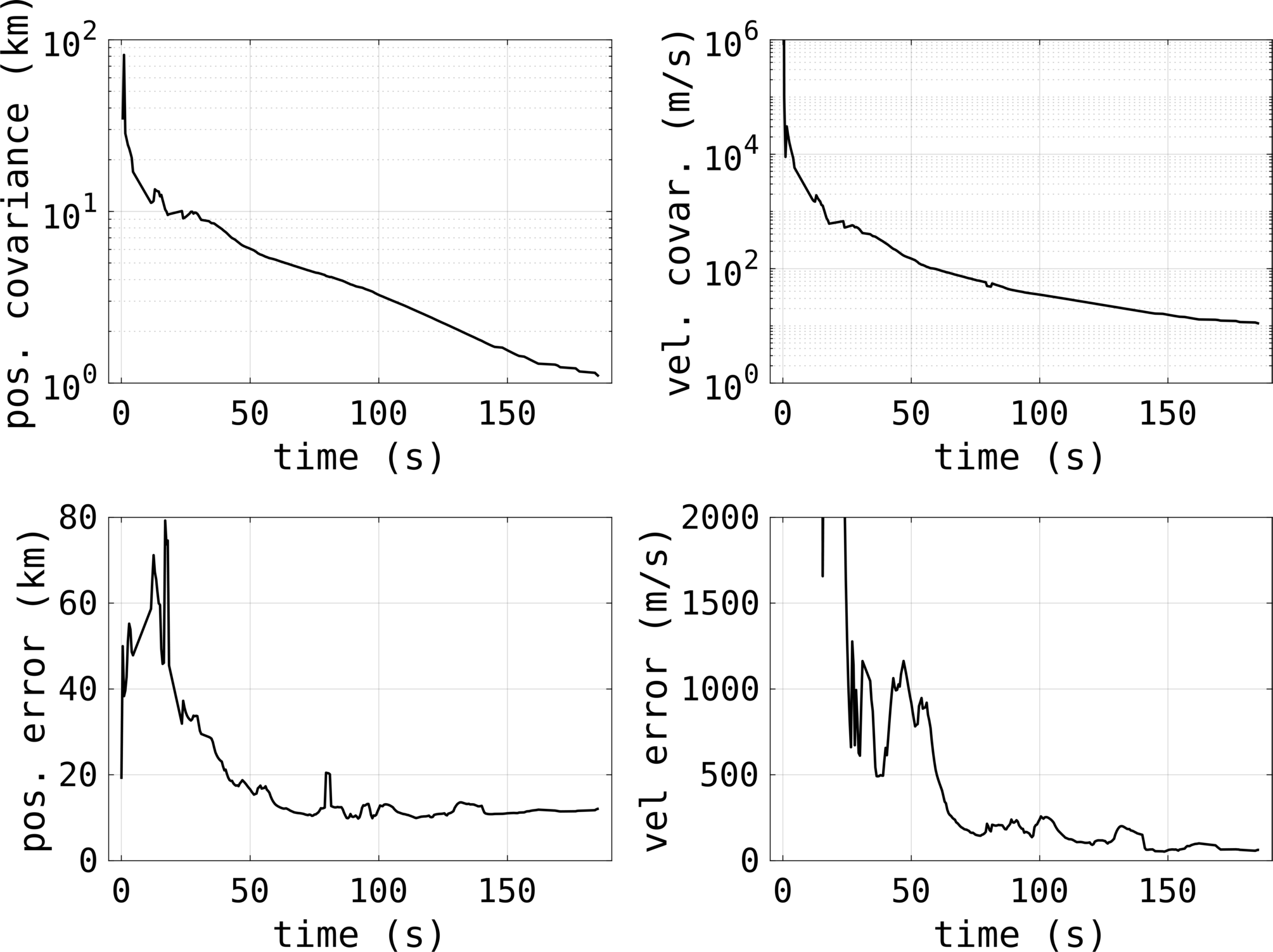}
\end{center}
\caption{The errors, and covariances, resulting from the OD process for a three minute track of the satellite OSCAR 31 (NORAD 19420).}
\label{fig:cool_result_NORAD_19420}
\end{figure}

Figure \ref{fig:cool_result_NORAD_19420} shows the orbit determination track covariances (position and velocity) as well as the accuracy against the TLE, and shows how all the measurements improve with the track length. This is a very typical result, in terms of accuracy, of a single pass of an RSO. After roughly one minute, a reasonably accurate position is often achieved. However, because of the coarse range resolution and spatial resolution the positional accuracy plateaus. The final orbital estimate, from a track approximately three minutes long, had a mean positional error of 17~km and a mean velocity error of 27~m/s. The velocity estimate is the most important, as any positional estimate errors will be dwarfed very quickly when propagated forward if the velocity is not sufficiently accurate.

For a low-bandwidth radar system, with long-CIT processing, the Doppler parameter will be a radar's most accurate, and normally would lead to greatly improve orbital accuracy~\cite{hennessy2022establishing}. However, as described in Section \ref{sec:sec4}, due to the impact of the ionosphere, the Doppler is biased and cannot be used to the extent that its accuracy would ordinarily allow.  Instead, the results shown here are primarily derived from the (notionally less accurate) range and spatial data, rather than Doppler.

The orbits from very short tracks can be quite inaccurate, and typically a track would need to be at least a minute long for it to be useful (on its own). For the detections shown in Figure \ref{fig:one_hour_detections}, for the tracks that have an associated TLE and are longer than a minute, the median positional errors are 28 km, and the median velocity error is 182 m/s. These errors reduce if the track length threshold is raised. The number of objects with no associated TLE is consistent with other results~\cite{vierinen20192018}. However, these observations were near sunset and so observations are dominated by satellites in a sun-synchronous orbit~\cite{Vallado2001fundamentals}.

\subsection{Initial Orbit Determination}\label{ssec:iod}

An initial orbit estimate is required for the numerical non-linear batch least-squares estimation. However, forming an initial orbit also acts as a {sanity check} that the track does indeed match an object in an Earth-centered orbit. Any attempt to form an orbital estimate from a noise, or other, track will not produce sensible results. Also, if the ambiguous range (the $J$ term in \eqref{eq:measured_bistatic_range}) has not been corrected beforehand, then it is immediately obvious at the output of the IOD function, and can be corrected here. This discrimination allows the system to operate with a higher WRF for a greater unambiguous Doppler span, as the range ambiguity can be corrected here.

The IOD stage for the HFLOS tracks uses the Herrick--Gibbs method~\cite{escobal1965methods}, requiring three positions at three corresponding times to provide an estimate for the velocity at the middle position. The Herrick--Gibbs approach is the result of a polynomial approximation to the position of the RSO over time, assuming a two-body orbital model, and is well suited for this problem as it is very accurate for short-arc measurements and is less sensitive to measurement errors than other approaches.

With three position measurements $\boldsymbol{r}_1$, $\boldsymbol{r}_2$, and $\boldsymbol{r}_3$, along with the respective time of flights between all three, $t_{21}$, $t_{32}$, and $t_{31}$, the velocity at the middle point, ${\boldsymbol{v}}_2$, is given by:
\begin{IEEEeqnarray}{rCL}
    \label{eq:herrick_gibbs}
    {\boldsymbol{v}}_2  &=& -t_{32}\left(\frac{1}{t_{21}t_{31}} + \frac{\mu}{12{\lvert \boldsymbol{r}_1\rvert}^3}\right)\boldsymbol{r}_1 + \nonumber \\ 
     && ~~~~~(t_{32} - t_{21})\left(\frac{1}{t_{21}t_{32}} + \frac{\mu}{12{\lvert \boldsymbol{r}_2\rvert}^3}\right)\boldsymbol{r}_2 +\nonumber \\
& & ~~~~~~~~~~~~~~t_{21}\left(\frac{1}{t_{32}t_{31}} + \frac{\mu}{12{\lvert \boldsymbol{r}_3\rvert}^3}\right)\boldsymbol{r}_3~.
\end{IEEEeqnarray}

This position and velocity, $\boldsymbol{r}_2$ and ${\boldsymbol{v}}_2 $, define a full orbit, and are the IOD estimate for the RSO. The estimate is improved with a greater time between estimates, so the first detection, a middle detection, and the final detection for the track are used for the three positions.

\subsection{Linear Array Initial Orbit Determination}
Successful IOD with detections from a linear receive array is more difficult. A two-dimensional array is able to discriminate in the two dimensions, and so provide a fixed spatial estimate with each detection. That is, a detection is able to specify a three-dimensional position, with an accuracy dependant on the range resolution and the size of the array. However, a linear array only senses in one dimension, and so any detection can only define a circle of potential detection locations. Because a detection does not define a position, it is challenging to estimate a target's trajectory along a track.

Earlier work used two detections, at the start and the end of a track, as well as a circular orbit assumption, to perform IOD~\cite{frazer2013orbit}. This approach results in two IOD estimates. However, with knowledge of the transmit beam pattern one of the IOD estimates is able to be discounted.

Given a detection from a linear receive array with range $\rho_{r}$ and coning angle $\alpha$, assuming a monostatic/quasi-monostatic configuration, the detection defines the position of the RSO as a circle, which is parameterised by the arbitrary angle $\psi$ such that:
\begin{IEEEeqnarray}{rCL}
    \boldsymbol{r}(\psi) = \boldsymbol{r}_{r} + \rho_{r}\left( \cos{\alpha} \boldsymbol{\mu}_p + \sin{\alpha}\left( \sin{\psi}\boldsymbol{b}_1 + \cos{\psi}\boldsymbol{b}_2 \right) \right),~~~~~\label{eq:circle_work}
\end{IEEEeqnarray}  where $\boldsymbol{b}_1$ and $\boldsymbol{b}_2$ are arbitrary principle axes of the circle, typically chosen such that $\boldsymbol{b}_2$ directs radially outward from Earth (assuming the array is horizontally flat and $\boldsymbol{\mu}_p$ lies in the north-east plane, $\boldsymbol{b}_1 = \frac{\rhovec_{r}}{\rho_{r}} \times \boldsymbol{\mu}_p$ and $\boldsymbol{b}_2 = \boldsymbol{\mu}_p \times \boldsymbol{b}_1$).

Instead of using a circular orbit assumption, given two detections, all possible angles can be searched through to find the most suitable solution for the orbit. Two positions (and the time of flight between them) can be used to determine an orbit. {Determining the orbit from such inputs is known as Lambert's problem~}\cite{escobal1965methods,izzo2015revisiting}. The two hypothesised angles, $\psi_1$ and $\psi_2$, define two positions (per \eqref{eq:circle_work}), $\boldsymbol{r}_1(\psi_1)$ and $\boldsymbol{r}_2(\psi_2)$. These two positions, and the time of flight, can be used to determine (utilising any method~\cite{izzo2015revisiting}) the orbit, and so the velocities at each point. These orbits can be verified to produce sensible velocities before continuing. For example, orbits with nonsensical eccentricities, such as being too fast to be Earth-centered, or with perigees too low, can be discounted. Additionally, the velocities can be checked to ensure they correspond with the respective Doppler measurements. The intersection of these limitations form tightly constrained admissible regions at which the true orbit must exist.

The vast majority of orbits from each angle, $\psi_1$ and $\psi_2$, can be discounted, especially with the Doppler constraints, and the remaining possibilities are used as the IOD estimate as part of the measurement-fitting OD processing. The orbit that results in the smallest residuals between itself and the measurements is deemed the correct solution.

Although there are many methods to determine a solution, for assessing a large number of orbits (from a very short arc), the Herrick--Gibbs approach can be used to approximate a solution to Lambert's problem. If $\boldsymbol{r}_1$ and $\boldsymbol{r}_2$ are the two positions, with a time-of-flight between them of $t$, a polynomial approximation to the position results in the velocities (and so orbits) at points $\boldsymbol{r}_1$ and $\boldsymbol{r}_2$ being given by $\boldsymbol{v}_1 = \frac{\boldsymbol{r}_2 - \boldsymbol{r}_1}{t} + \frac{\mu t}{6}\left( \frac{\boldsymbol{r}_1}{{{\lvert \boldsymbol{r}_1 \rvert}^3}} - \frac{2\boldsymbol{r}_2}{{{\lvert \boldsymbol{r}_2 \rvert}^3}} \right)$ and $\boldsymbol{v}_2 = \frac{\boldsymbol{r}_2 - \boldsymbol{r}_1}{t} - \frac{\mu t}{6}\left( \frac{2\boldsymbol{r}_1}{{{\lvert \boldsymbol{r}_1 \rvert}^3}} - \frac{\boldsymbol{r}_2}{{{\lvert \boldsymbol{r}_2 \rvert}^3}} \right)$, respectively.

\begin{figure}[ht]
\begin{center}
    \includegraphics[width=\columnwidth]{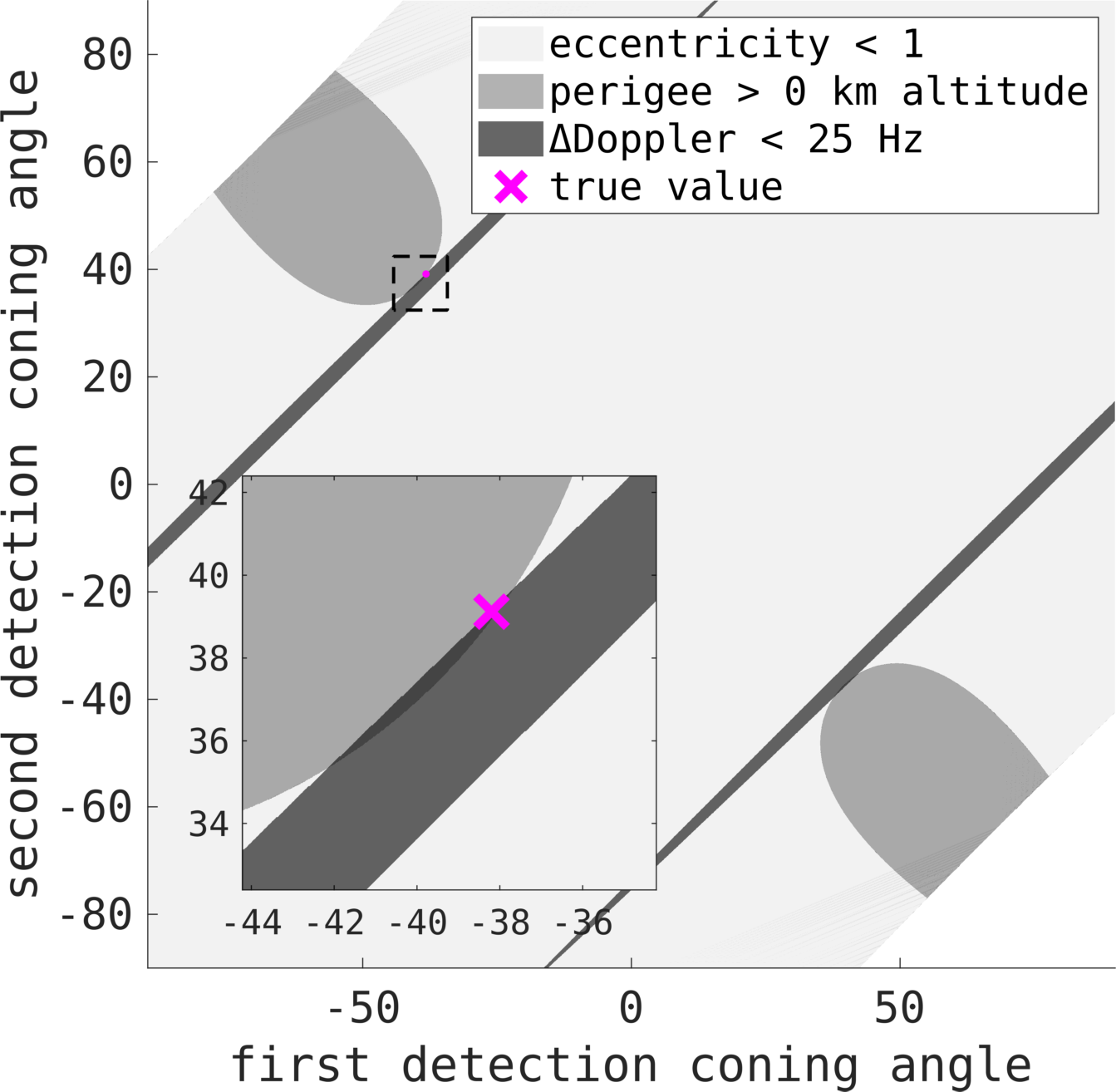}
\end{center}
\caption{Admissible regions for the IOD solution from the first detection and last detection of the pass of the satellite HJ-1C (NORAD 38997). The orbits are defined by the detections' circle angles $\psi_1$ and $\psi_2$ and the three regions are from the valid eccentricity, valid perigee and matching Doppler. There are two very small sections defined by the intersection of all the regions, an inset highlights one of them along with the satellite's true orbit.}
\label{fig:admissible_regions}
\end{figure}

Figure \ref{fig:admissible_regions} shows the admissible regions from the first and last detections of a 95-second pass of the synthetic aperture radar satellite HJ-1C (NORAD 38997). A maximum Doppler difference of 25 Hz has been chosen to (generously) ensure a valid solution is found despite any errors in range, Doppler, and coning angle, either from coarse measurement resolution or the impact of ionospheric propagation. The three regions are valid eccentricity, sensible perigee, and matching Doppler, and the narrow intersections of these three regions rule out the vast majority of the potential orbital parameter space. Much like the circular assumption, these admissible regions will generally produce two valid IOD solutions (as shown by the two small intersection regions in Figure \ref{fig:admissible_regions}). However, the incorrect solution results in a worse measurement fit, in terms of the residuals, and can be discounted. Other admissible regions, such as transmit and receiver antenna beampatterns, or other orbital regimes of interest, can also be applied.

For the HJ-1C pass, after searching through all realisable IOD options, the resulting orbit that best fit the measurements had a mean positional error of 12~km and a mean velocity error of 160~m/s when compared with the TLE. For linear array measurements, the velocity error tends to be higher than orbits formed from two-dimensional detections. However, the errors in positional estimates are typically comparable.

A similar approach has used a particle filter to sample the large number of potential orbits matching the initial and final range and coning angle measurements to produce IOD estimates~\cite{white2024australian}.

\subsection{Multi-pass Orbit Determination}\label{ssec:multipass}
For a single pass, the radar only observes the RSO for a tiny fraction of a full orbit, which can make accurate orbital estimates challenging. Better results can be achieved from using multiple passes for the OD step, as the greater time span assists in more precisely determining the velocity. However, the biggest challenge with multi-pass orbit determination is that the simple two-body (or Keplerian) orbital model will not suffice. Aspects such as a non-spherical Earth, the impact of other bodies in the solar system, and drag caused by the upper atmosphere all contribute to an RSO's orbit. These factors need to be taken into account for future prediction and multi-pass propagation.  The software package Orekit is used to propagate orbits whilst taking into account the non-Keplerian aspects for accurate multi-pass propagation~\cite{maisonobe2010orekit}.

Given tracks from two passes of one RSO, along with each track's associated detections and orbit, the combined OD process is {near identical} to the single-track OD, only now the measurement function $\boldsymbol{f}$ is far more complicated {due to the inclusion of the other forces operating on the RSO} and cannot be evaluated analytically. {A single position estimate from each pass is used to numerically determine an IOD estimate by solving Lambert's problem utilising the more detailed propagation, which takes} into account upper atmospheric drag, oblate Earth, and other gravitational sources. Like the single-pass OD step, this IOD estimate is used as the initial solution for the batch least-squares method, {also} with the new propagator.

With the time span between two tracks being hours, or even days, it is easier to more tightly constrain the velocity and produce better orbital estimates.

\begin{figure}[ht]
\begin{center}
    \includegraphics[width=\columnwidth]{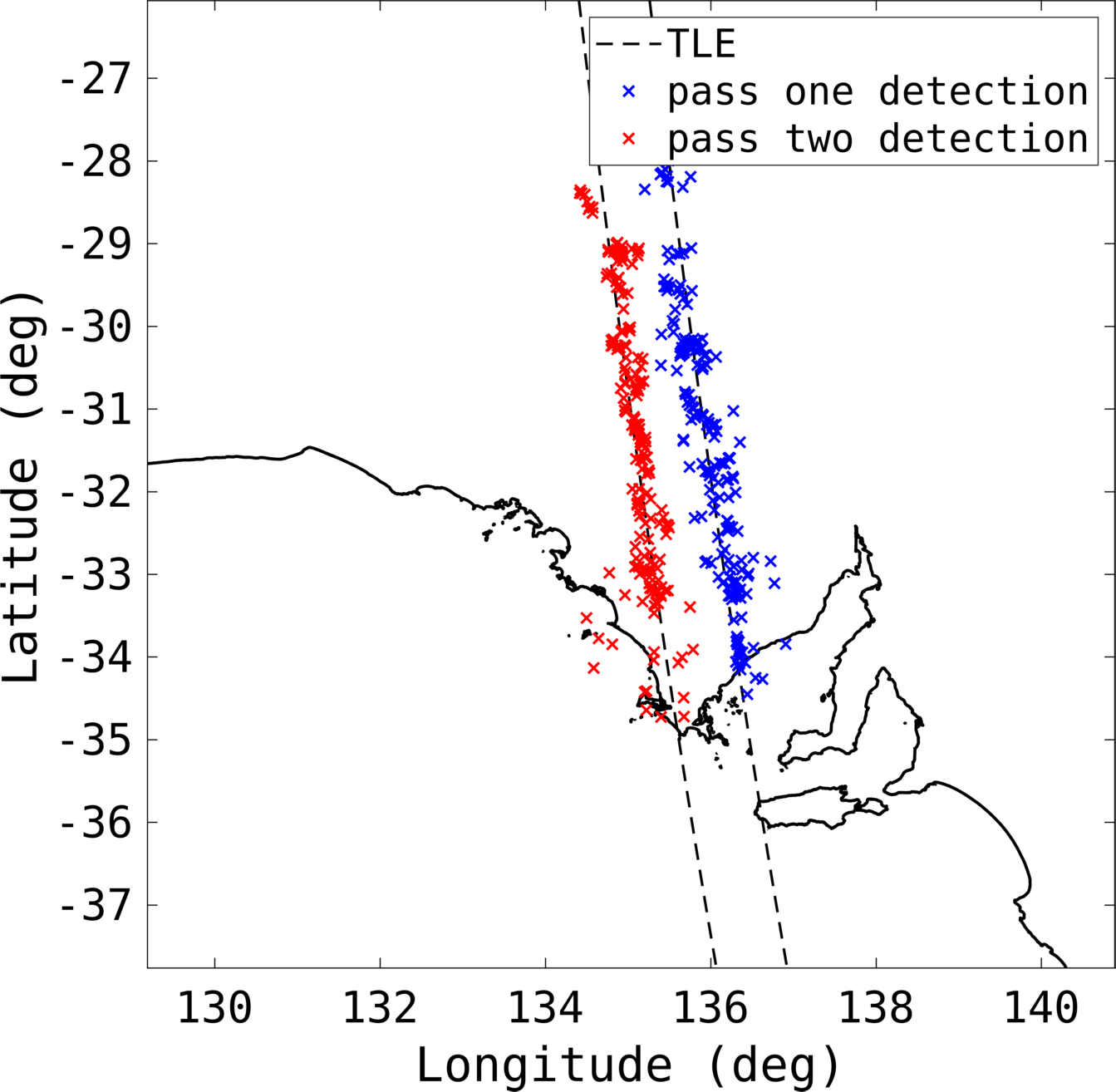}
\end{center}
\caption{The detections, as well as the TLE, of two passes of a rocket body (NORAD 6393) are shown.}
\label{fig:SUPER_multipass_orbit_map1}
\end{figure}

Figure \ref{fig:SUPER_multipass_orbit_map1} shows the detections from two subsequent passes, roughly 24~hours apart, of a rocket body (NORAD 6393) above South Australia.  The detections line up well with the TLE. However, the impact of the coarse measurement parameters is evident, especially with the lower elevation detections at the southern end of the tracks.

\begin{figure}[ht]
\begin{center}
\includegraphics[width=\columnwidth]{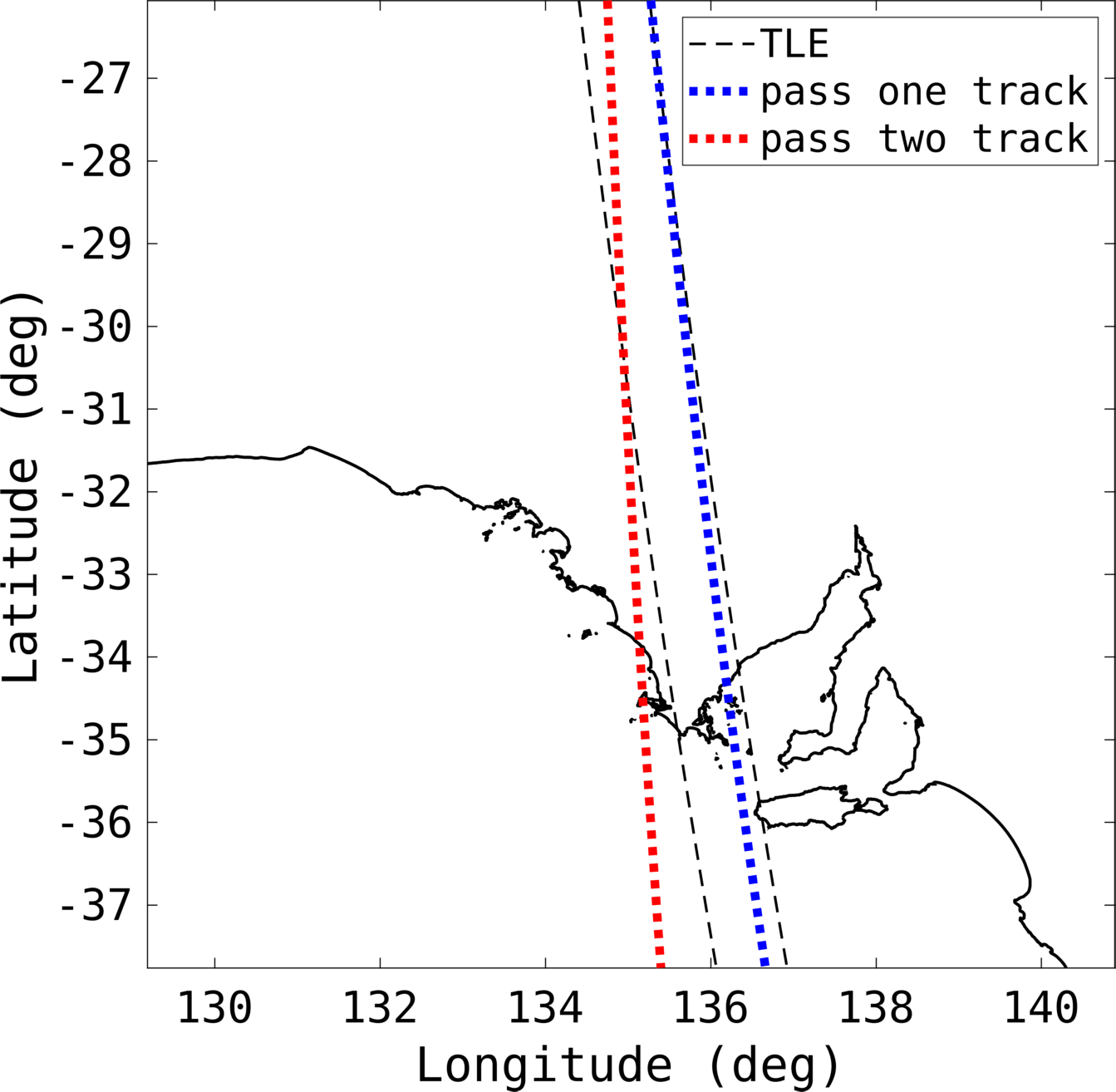}
\end{center}
\caption{The determined orbits, as well as the TLE, of two passes of a rocket body (NORAD 6393), corresponding to the detections in Figure \ref{fig:SUPER_multipass_orbit_map1} are shown.}
\label{fig:SUPER_multipass_orbit_map2}
\end{figure}

Figure \ref{fig:SUPER_multipass_orbit_map2} contains the same passes as Figure \ref{fig:SUPER_multipass_orbit_map1} but now displaying the two orbits resulting from the two tracks. {The OD process applied independently to each track results in mean positional errors of less than 20~km for both; however, the mean velocity errors are 228~m/s and 264~m/s, respectively, which is not accurate enough for proper propagation.} This is particularly evident in the track from the second pass, as it diverges from the TLE rapidly. Despite this, the tracks were able to be associated with the same object without requiring the TLE.

\begin{figure}[ht]
\begin{center}
\includegraphics[width=\columnwidth]{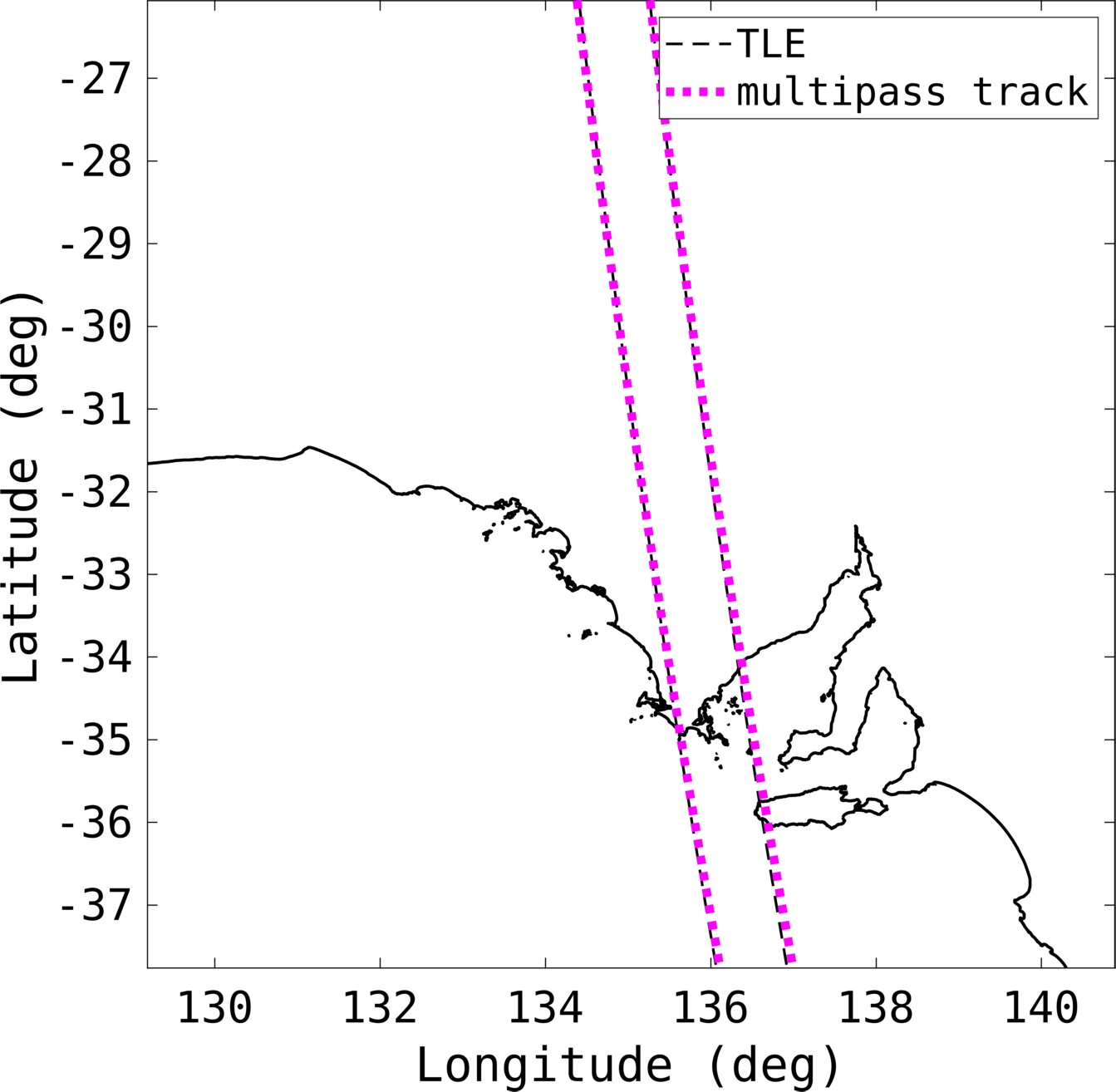 }
\end{center}
\caption{The resulting orbit, as well as the TLE, of two passes of a rocket body (NORAD 6393), corresponding to the detections in Figure \ref{fig:SUPER_multipass_orbit_map1} and the tracks in Figure \ref{fig:SUPER_multipass_orbit_map2}, are shown.}
\label{fig:SUPER_multipass_orbit_map3}
\end{figure}

Figure \ref{fig:SUPER_multipass_orbit_map3} shows the result of the full multi-pass orbit from the detections and tracks in Figures \ref{fig:SUPER_multipass_orbit_map1} and \ref{fig:SUPER_multipass_orbit_map2}, respectively. The resulting orbit is far more accurate, with a mean positional error of 9~km and a mean velocity error of 19~m/s. The positional errors are approaching the inherent uncertainty of the TLE itself, and the velocity error is small enough such that the track can be propagated forward accurately enough to associate future passes and maintain an ongoing catalogue.

The key to accurate predictions, namely overcoming some of the coarse detection accuracy, is combining multiple tracks, hours or days apart, to accurately constrain the orbit.

\subsection{Linear Array Multi-pass Orbit Determination}
Multiple tracks from linear array observations can also be used to determine very accurate orbital results. Like the general planar array approach, the two positions can be used to form a solution to Lambert's problem with an accurate propagator, and this can be used to find a solution with the least-squares approach. The dominant challenge with the linear array tracks, however, is that if there is not a TLE association, the individual tracks' velocity estimates are not always accurate enough to identify future passes.  

It is possible to achieve good results by attempting {a brute-force search} of unassociated tracks; however, {unlike the two-dimensional array case, this search space cannot be easily prune, and the large number of permutations quickly makes this approach impractical}. Robust IOD and OD with tracks from linear arrays is an ongoing area of work.

\subsection{Real-time Mid-pass Cueing}

During these deployments, space surveillance sensors were networked and HFLOS detection and track information was disseminated. These data were used for cueing narrow-field-of-view sensors with tracks of RSOs that had no associated TLE. One of the sensors was an EO telescope system, described in Appendix \ref{sec:appendix_telescope}.

\begin{figure}[ht]
\begin{center}
\includegraphics[width=\columnwidth]{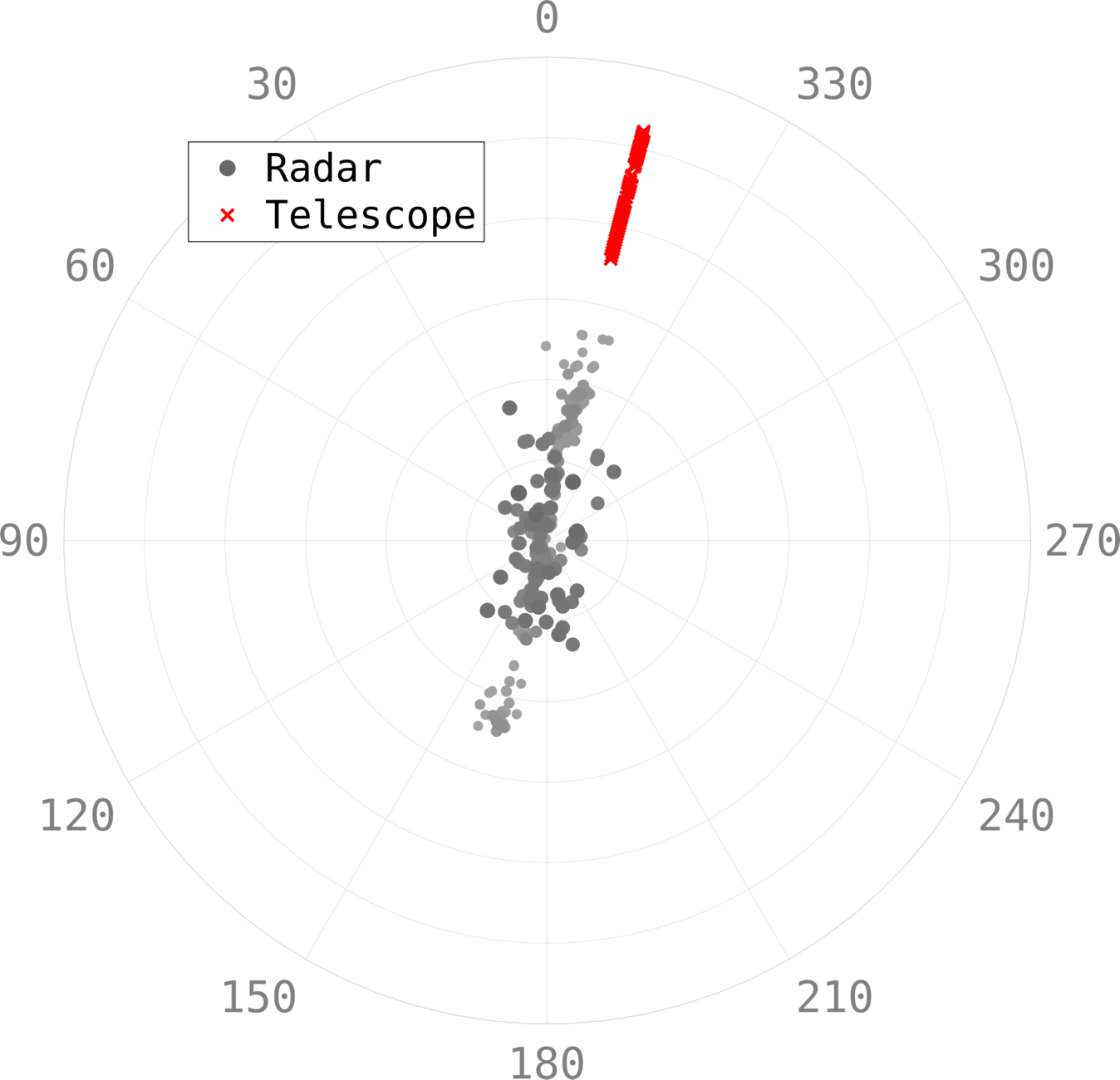}
\end{center}
\caption{Radar and EO telescope spatial detections of a pass of an RSO are shown. The telescope was slewed to an orbit formed by the radar observations.}
\label{fig:telescope_cueing}
\end{figure} 

Figure \ref{fig:telescope_cueing} shows the detections of a satellite pass from the HFLOS radar and also one from the DSTG EO telescope. The telescope was cued with an orbit formed from the radar detections during the pass. Because of the telescope software's interface, a pseudo-TLE had to be formed from the HFLOS-determined orbit. This was challenging because the track had to be formed and track handoff needed to occur before the RSO moved into the Earth's shadow, or below the horizon. Despite the relatively coarse spatial accuracy of the radar (highlighted in the distribution of detections in the figure), the resulting orbit was sufficiently accurate that the narrow-field telescope system was able to acquire the track and continue following the RSO for a further 158 seconds to low elevations.

\begin{figure}[ht]
\begin{center}
\includegraphics[width=\columnwidth]{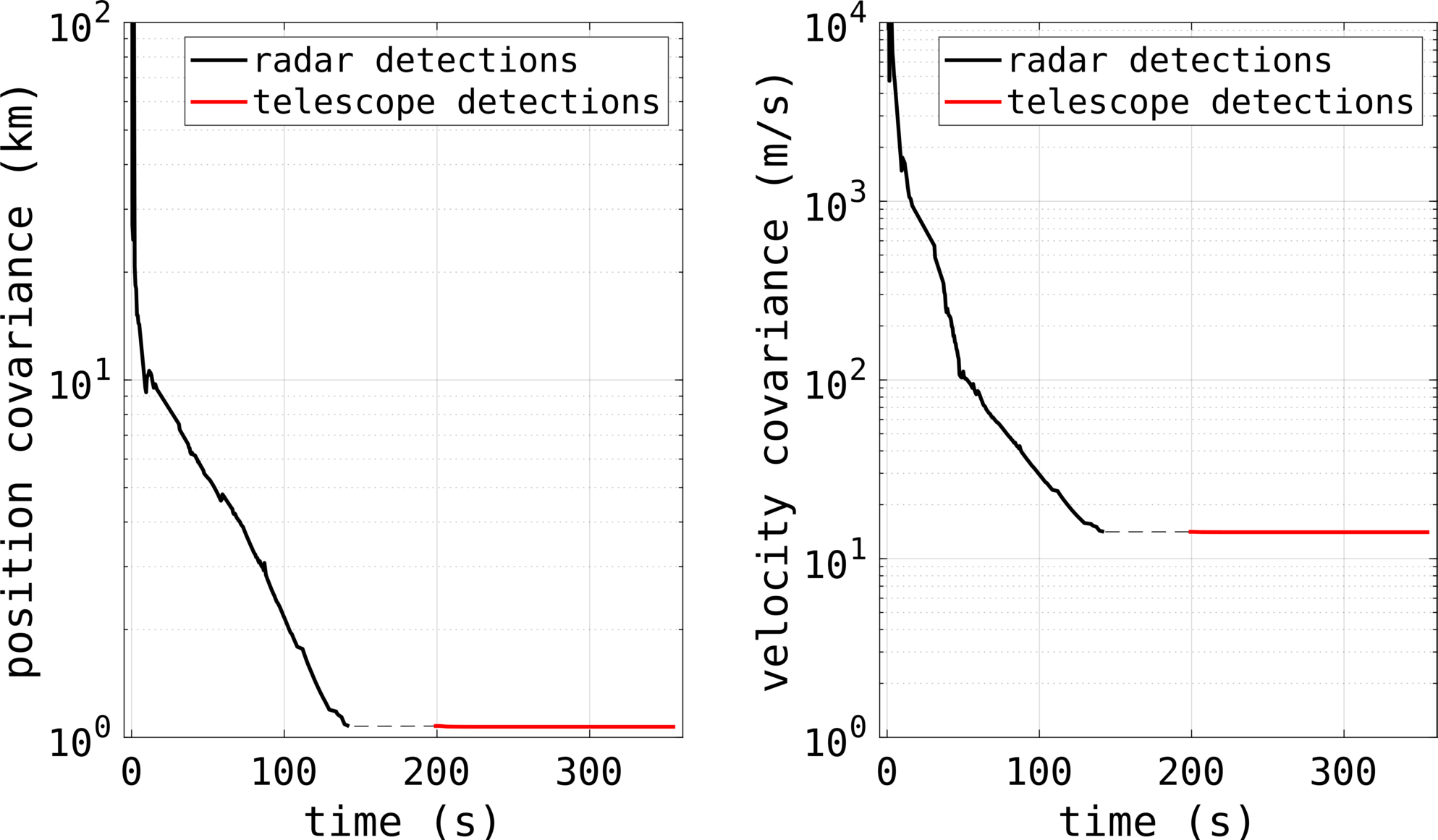}
\end{center}
\caption{The covariances of the orbit formed with both the radar and EO telescope detections.}
\label{fig:radar_telescope_fused}
\end{figure}

This track had no associated TLE. However, the detections from both sensors were fused to produce a combined orbit. The covariances of the fused orbit are shown in \mbox{Figure \ref{fig:radar_telescope_fused}} and indicate good confidence in the track's accuracy, as well as the fact that the telescope was able to observe the RSO for a further 158 seconds. This highlights the benefit of the real-time signal processing capabilities in that accurate track information can be formed immediately.

\section{Conclusion} \label{sec:sec7}
This paper has described the use of HF radar for the surveillance of space, highlighting the advantages in the detection and tracking of large objects in low-to-medium Earth orbit, whilst also detailing the challenges that come from using such a low frequency. Results from deployments of DSTG's HFLOS radar have been described, as well as a brief overview of the radar product formation, signal processing, detection, ionospheric correction, and tracking stages. HF radar is well suited to space surveillance, with a small deployable system providing real-time persistent wide-area surveillance for the detection and tracking of RSOs (both catalogued and uncatalogued). By combining multiple tracks, it is possible to generate orbit estimates that are sufficiently accurate for maintaining custody of tracks and providing space domain awareness.

\section{Acknowledgements}
The work described here is the result of many people's efforts over many years, unfortunately too many to name here. However, the authors would like to specifically acknowledge, and thank, Gordon Frazer, Travis Bessell, Mark Rutten, Stephen Howard, Nicholas Moretti, Kruger White, Chris Walden, Gavin Scarman, and Mark Jessop. The authors would also like to acknowledge the broad support of the HF Systems and HF Sensing \& Protection groups at DSTG.
Some of the results published in this paper were obtained using the HF propagation toolbox, PHaRLAP, created by Manuel Cervera, Defence Science and Technology Group, Australia (manuel.cervera@dsto.defence.gov.au). This toolbox is available by request 
from its author.

\bibliographystyle{IEEEtran}
\bibliography{mainbib}

\balance 
\begin{appendices}

\section{DSTG Tracking Telescope}\label{sec:appendix_telescope}

DSTG operates several robotic EO telescope systems, of various sizes, for space surveillance~\cite{eastment2014technical,hobson2016catalogue,moretti2017autonomous,white2024australian}. These systems have been used for surveillance of orbital regimes from LEO out to beyond geosynchronous equatorial orbit (GEO), and have been involved in international space surveillance trials. For the SpaceFest activities, a portable system was transported to Far North South Australia. The system was equipped with an RH200 f/3 telescope and an FLI PL4710 cooled CCD camera, providing a 1.27\textdegree~field of view and an instantaneous field of view of 4.46 arcseconds per pixel. It was mounted on a Paramount ME2 equatorial mount and controlled using TheSkyX commercial software. 

The telescope can observe RSOs at night, provided they are not in Earth's shadow. Objects in LEO are typically visible for only one to two hours after sunset and before sunrise, whereas GEO objects remain visible for most or all of the night.

\begin{figure}[ht]
\begin{center}
\includegraphics[width=\columnwidth]{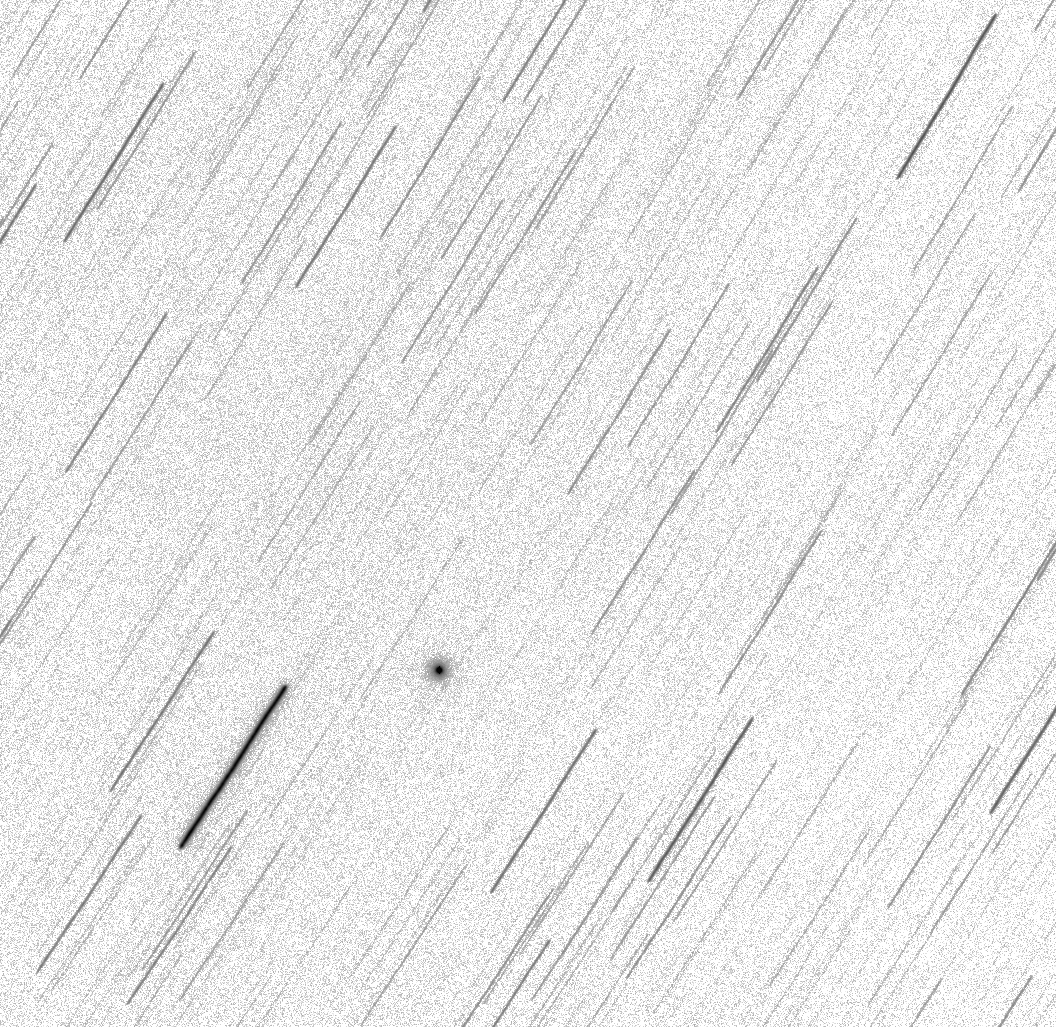}
\end{center}
\caption{Example image taken by the DSTG EO telescope system, with the telescope tracking an RSO in LEO. The RSO appears as a dot against a background of stars, appearing as streaks due to the telescope's motion. The colours have been inverted. }
\label{fig:telescope_example}
\end{figure}

The system is calibrated using an automated procedure, by collecting many star fields to build a TPoint model~\cite{wallace1994tpoint}. After calibration, the system is able to accurately point the telescope and collect data based on any provided TLE sets. The mount typically requires between 10~s and 30~s to slew and stabilise on a new LEO target. When the system has been slewed to a track, the telescope takes long exposure images. A GPS timing card is connected to the camera shutter to precisely time-stamp the opening and closing of the camera shutter.

Figure \ref{fig:telescope_example} is an example image captured by the telescope. The telescope's motion following the track results in stars appearing as streaks throughout the image. However, the RSO appears as a fixed dot. Long-exposure images help provide better sensitivity against faint objects.
\end{appendices}
\newpage

\begin{figure*}[ht]
\begin{center}
\vspace{-6ex}
\includegraphics[width=1.45\textwidth, angle=-90]{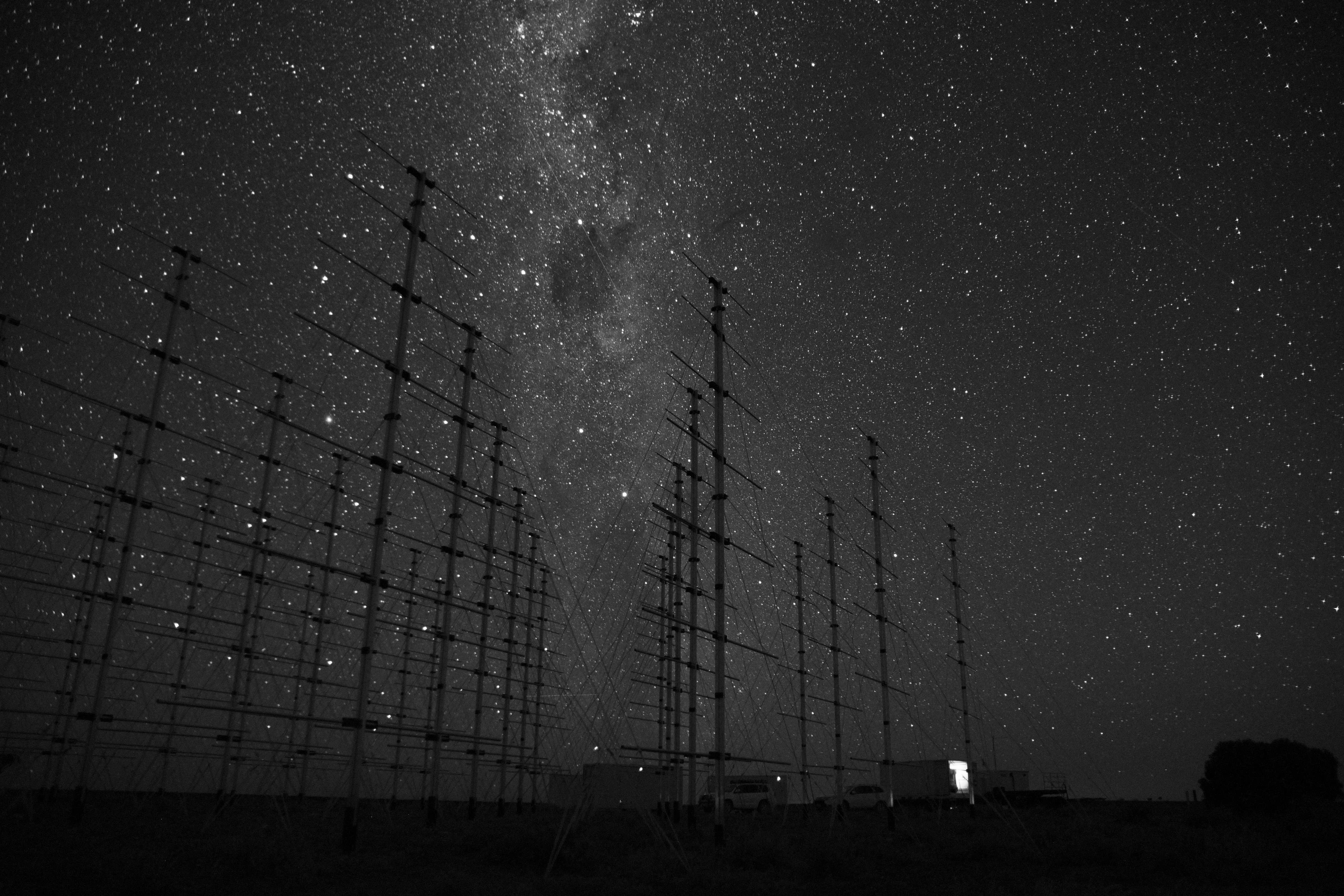}
\end{center}
\end{figure*} 
\end{document}